# Understanding and formalization of the fretting- wear behavior of a cobalt-based alloy at high temperature

Alixe Dreano*, Siegfried Fouvry*, Gaylord Guillonneau

University of Lyon, Ecole Centrale de Lyon, LTDS UMR 5513, Ecully, France

Corresponding authors : E-mail addresses: alixe.dreano@ec-lyon.fr (A. Dreano), siegfried.fouvry@ec-lyon.fr (S. Fouvry).

Abstract

The purpose of this study is to investigate the mechanisms involved in the wear of cobalt-based interfaces at high temperature. The studied contact is a cobalt-based alloy subjected to fretting against an alumina sample. At high temperature, a protective third body is spontaneously created at the interface and presents excellent tribological properties. The formation of the so-called "glaze layer" leads to an absence of wear. The investigation presents on complete description of the high-temperature tribolayer with microstructural, chemical and mechanical characterizations. The glaze layer regime is mainly related to a threshold temperature above which a thin cobalt-rich layer is formed by a tribo-sintering process. A formalization of the tribo-sintering process is proposed to predict the necessary number of fretting cycles $N_{GL}$ to form the glaze layer. The tribo-sintering process prevents wear debris ejection by continuously re-incorporating the wear debris particles in the glaze layer. The re-incorporation of the wear debris may be the reason for the absence of wear of the fretted interface from a macroscopic point of view. Finally, the paper presents an extended friction energy wear approach taking into account tribo-oxidation and tribo-sintering considerations. The formulation is able to predict wear for a large range of tribological parameters (temperature, frequency, sliding amplitude, number of cycles), when applied to the Co-based/alumina contact.

Nomenclature

| Material | |
|---|---|
| CoRL | Cobalt-Rich Layer |
| CRL | Chromium-Rich Layer |
| HS25 | Haynes 25, cobalt-based alloy |
| MOL | Mixed Oxide Layer |
| Loading conditions | |
| f | Frequency (Hz) |
| L, $L_{GL}$ | Sliding distance, sliding distance needed for the glaze layer formation (m) |
| N | Number of fretting cycles |
| $N_{eff}$ | Effective number of fretting cycles |
| $N_{GL}$ | Number of cycle needed for the glaze layer creation |
| P | Normal Force (N) |
| Q | Tangential Force (N) |
| $Q^*$ | Maximum tangential force (N) |

| T | Temperature (°C or K) |
|---|---|
| δ | Fretting displacement (mm) |
| $\delta^*$ | Displacement amplitude (mm) |
| $\delta_0$ | Sliding amplitude (mm) |
| µ | Conventional friction coefficient |
| $µ_e$ | Energetic friction coefficient |
| $\overline{µ_e}$ | Mean energetic friction coefficient |
| **Wear** | |
| V | Total wear volume (mm$^3$) |
| $\overline{V_{HS25}}$ | HS25 missing volume (mm$^3$) |
| $V_\phi$ | Offset of wear (mm$^3$) |
| $\Phi_F, \Phi_E$ | Debris formation flow, debris ejection flow |
| **Wear modelling** | |
| | |
| $E_a$ | Activation energy for the oxidation process (J/mol) |
| $E_d$ | Dissipated energy or friction energy (J) |
| $\Sigma E_d$ | Cumulated dissipated energy (J) |
| K | Archard wear coefficient (mm$^3$/J) |
| $K_{ox}$ | Oxidational wear coefficient (mm²/s$^{0.5}$/J or mm/s$^{0.5}$/N) |
| ΣW | Archard's work (J) |
| α | Energetic wear coefficient (mm$^3$/J) |
| $α_{ox}$ | Oxidational energetic wear coefficient (mm²/s$^{0.5}$/J or mm/s$^{0.5}$/N) |
| $α^*$ | Apparent energetic wear coefficient (mm$^3$/J) |
| Ψ | Tribo-oxidation parameter (mm.s$^{0.5}$) |
| **Tribo-sintering modelling** | |
| $D, D_0$ | Diffusion coefficient, diffusion constant (cm²/s) |
| $E_{a,GL}$ | Activation energy for the glaze layer formation (J/mol) |
| $E_f$ | Activation energy for the sintering process (J/mol) |
| k , m, n | Constants |
| r | Particles diameter (mm) |
| $S, S_{GL}$ | Sintering parameter, threshold value (s.mm$^{0.5}$) |
| $t, t_{GL}$ | Time, necessary time to form the glaze layer (s) |
| $λ, λ_{crit}$ | Interpenetration of two particles, minimum interpenetration require to form the glaze layer |
| **Subscripts** | |
| eff | Effective part of the considered variable |
| GL | Referred to the glaze layer |
| i | Related to the involved diffusion mechanism |
| ref | Related to the reference conditions |
| I,II,III | Referred to domains I, II and III |
| 1,2 | Referred to the transition temperatures |
| **Others** | |
| R | Universal constant (J/K/mol) |
| R² | Coefficient of determination |

Introduction

The changes in the wear behavior of metallic alloys when the temperature varies is a crucial aspect in many industrial components such as the turbine and the compressor components in aircraft engines. The blade/disk contact is subjected to cyclic thermal loading combined with cyclic sliding (fretting) which can significantly damage the interface (Fig. 1). Many works can be found in the literature dealing with this industrial issue. For example, Mary *et al.* studied the wear behavior at different temperatures of a Cu–Ni–In plasma coating/ Ti17 contact representative of the blade/disk contact in high pressure compressor [1]. Viat *et al.* [2] studied the high temperature wear behavior of a cobalt-based alloy/ceramic tribocouple representative of the blade/disk contact of turbines. These studies deeply examined the influence of temperature on wear but do not try to model such behavior. Hence, the understanding and the formalization of the temperature effect during fretting-wear is needed for such industrial issue. In the present paper, the wear behavior of a cobalt-based alloy, widely used for aeronautical components [2–4] is studied in a more fundamental way (simplified contact configuration and inert counterbody) in order to deepen the actual knowledge on the dependence of the Co-alloy wear mechanism with the temperature.

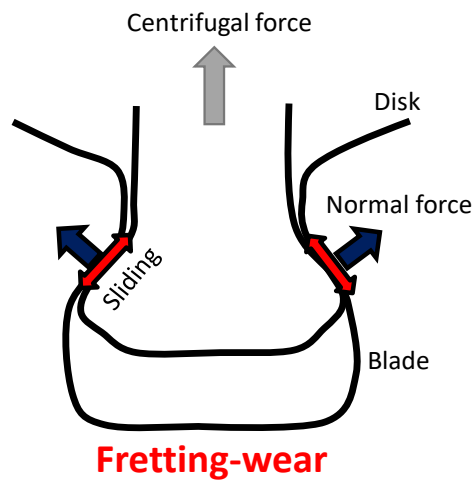

Fig. 1: Fretting-wear process at the blade/disk contact

The wear rate of cobalt-based alloys highly depends on temperature [1,5]. Above a certain transition temperature, wear changes from severe to mild. In the mild wear domain, a lubricant tribolayer is naturally created in the interface leading to a drastic reduction of the wear volume and the friction coefficient [6,7]. The so-called "glaze layer" is formed of oxidized and compacted/sintered wear debris generated during the transient regime [8,9]. According to Jiang *et al.* [9], the compaction of triboparticles is enhanced by elevated temperatures and small grain sizes. Kato *et al.* [10] called "tribo-sintering" the capacity of debris to sinter during sliding. They showed that mild wear is promoted by enhanced diffusion properties of the oxide debris particles. Hence, they demonstrated that the formation of the glaze layer can be related to a sintering process of the debris particles. Showing the significant influence of diffusion to form the protective glaze layer structure, they suggested that the wear volume is proportional to the sliding distance until the glaze layer is formed ($V \propto L_{GL}$). Viat *et al.* [11] confirmed that the formation of glaze layer on Haynes 25 is mainly controlled by cobalt oxides which presents the highest auto-diffusion coefficient among the main alloying elements (Co, Cr, Ni).

In the literature, it is shown that the glaze layer is mainly nanocristalline [12,13] with the presence of some amorphous zones near the fretted interface. Nanocrystalline materials present high strength and as high hardness, such as the glaze layer [14,15], explaining the use of nanocristalline structure for tribological applications since they better resist to wear than conventional alloys [16,17]. In addition, Viat *et al.* [15] proposed that the tribological properties of the glaze layer are mainly due to the perfect ductile behavior observed at high temperature whereas a brittle behavior is observed for low temperature.

At low temperature, when the agglomeration and compaction of the wear debris are not promoted, the wear rate and the friction coefficient are high. Recently, it was shown that the wear evolution in the severe wear domain is controlled by a synergistic action of the oxidation of the surface and the abrasion of the oxidized surface [18]. The oxidized layer is continuously ejected out of the interface leading to high abrasive wear rates. An extended energetic wear model was also proposed in order to capture the transition from severe to mild wear [19], when the tribological conditions are sufficient to generate the compaction and sintering of the powdered debris layer.

The objective of this paper is to investigate in a more fundamental way, rather than industrial purpose, the high temperature wear mechanisms of cobalt-based alloys subjected to fretting. First, an overview of the different wear processes occurring over a large range of temperatures (25°C – 600°C) is presented. Then, the investigation focuses on the high temperatures at which the formation of the glaze layer leads to an unworn regime. A microstructural, chemical and mechanical description of the glaze layer is proposed in order to completely describe this protective structure. It is then proposed to understand the protective tribolayer formation by considering a sintering formalism as previously intuited by Kato and co-authors [10,20]. The effectiveness of the glaze layer is also discussed by considering two hypotheses; one based on the mechanical behavior of the fretted interface and the other based on the sintering formalism. Finally, the paper extends an "effective" friction energy wear approach taking into account the tribo-oxidation and sintering processes activated at the fretted interface. The proposed wear model is then able to predict the wear volume for a cobalt-based alloy/alumina contact irrespective of the operating temperature.

1. Experimental procedure

    1.1. Materials

The specimens used in this study are a cobalt-based alloy (HS25) and a ceramic (alumina). Cobalt-based alloys are extensively used in the industry since they present good mechanical properties and corrosion resistance at high temperature [2,3]. Chromium, nickel, tungsten and other alloying elements, presented in Table 1, are present in solid-solution or in the form of carbides [18]. The mechanical properties of the Co-based alloy are given in Table 2. The main alloying element is chromium which provides corrosion resistance by creating a protective $Cr_2O_3$ layer at the surface [21,22]. The alumina sample was chosen for its mechanical resistance (Table 2) and its inertness in the wear process as it was demonstrated in previous studies [18,19]. Hence, the study will only focus on the intrinsic wear mechanism of the cobalt-based alloy HS25.

Table 1: Nominal composition of the Haynes 25 alloy (at. %) extracted from supplier documentation

| Co | Cr | Ni | W | Fe | Mn | Si | C |
|---|---|---|---|---|---|---|---|
| 54 | 24 | 11 | 5 | 3.3 | 1.6 | 0.7 | 0.4 |

Table 2: Mechanical properties of the studied materials (Haynes 25 and alumina) extracted from supplier documentation

| Material | Modulus of elasticity (GPa) | Poisson's coefficient | Yield strength (MPa) | Hardness (Hv) |
|---|---|---|---|---|
| Alumina | 310 | 0.27 | 2000 (resistance to compression at 20°C) | 1800 |
| HS25 | 225 (20°C), 181 (600°C) | 0.31 | 654 (25°C), 445 (400°C) | 251 (20°C), 171 (425°C) |

1.2. Fretting tests

The studied interface consists of a cross-cylinders contact configuration as presented in Fig. 2 a). The rods have a diameter of 8 mm leading to an equivalent sphere-on-flat contact.

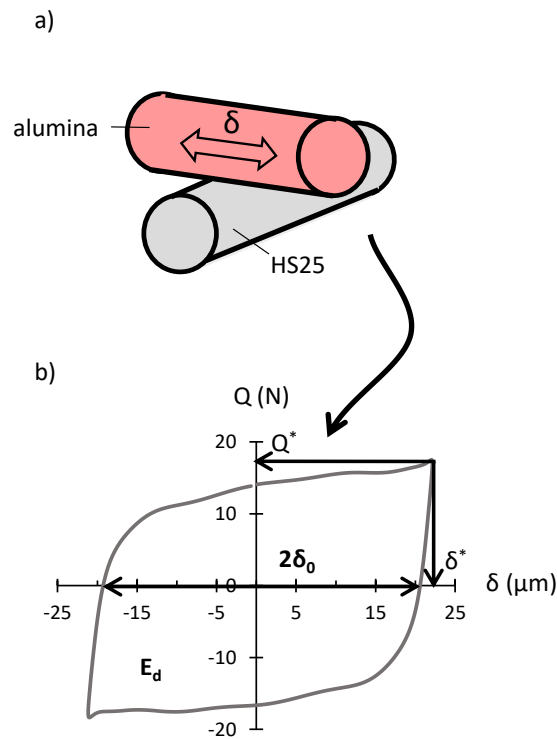

Fig. 2: a) Alumina/HS25 cross-cylinders configuration; b) Fretting loop (f = 50 Hz, P = 50 N, N = 200 000 cycles, $\delta_0$ = ± 20 µm, T = 575°C)

The high-temperature fretting device developed at the LTDS laboratory was presented previously in [18]. It is composed of a fixed part <u>rubbing against</u> a moving part which is set in motion by an electromagnetic shaker. The specimens are fixed to the sample holders and pressed with a normal force *P.* The normal force *P* is recorded along the test as well as the relative displacement $\delta$ and the tangential force *Q*. Fretting loops, showing the evolution of *Q* versus $\delta$, are plotted for each cycle as presented in Fig. 2 b). The sliding amplitude $\delta_0$ is then defined as the real displacement at the interface and measured when the tangential force $Q$ is equal to 0 (Fig. 2 b)). All the tests were conducted by monitoring the sliding amplitude $\delta_0$. The dissipated energy is finally defined as the area of the fretting loop:

$$E_d = \int_{-\delta^*}^{+\delta^*} Q(\delta)d\delta \approx 4Q\delta_0 \qquad (1)$$

### 1.3. Loading conditions

To go further in the wear analysis, a cross-strategy was adopted and centered on a reference test condition: $f_{ref}$ = 50 Hz, $\delta_{0,ref}$ = ± 20 µm and $N_{ref}$ = 200 000 cycles. The reference temperature depended on the wear mechanism involved in the contact: $T_{I,ref}$ = 100°C for the low temperature domain, $T_{II,ref}$ = 230°C for the medium temperature domain and finally $T_{III,ref}$ = 575°C for the high temperature domain. In addition, all the parameters varied between the following ranges: f = [1 ; 70] Hz, $\delta_0$ = [±5 ; ±160] µm, N = [0 ; 300 000] cycles and finally T = [25 ; 600]°C. The normal force was fixed to 50 N. The initial maximum hertzian contact pressure was equal to 2.3 GPa which generated plastic deformations of the HS25 sample in the beginning of the test [18].

During the test, the friction coefficient was computed using Eq. (2).

$$\mu = \frac{Q^*}{P} \qquad (2)$$

In this study, the energetic friction coefficient was also considered and was defined as follows:

$$\mu_e = \frac{E_d}{4P\delta_0} \qquad (3)$$

### 1.4. Fretting wear scar analysis

#### 1.4.1. Wear volume measurements

After the tribological tests, the samples were cleaned in an ultrasonic bath with ethanol in order to remove the loose debris particles. Fretting scars were optically observed and topography scans were performed and analyzed using an interferometric BRUKER system (Fig. 3 a)). Fig. 3 b) shows the wear tracks of alumina and HS25 after a numerical subtraction of the cylinder shape and Fig. 3 c) presents the equivalent 2D profiles of the fretted interface. It was shown that the alumina sample did not undergo wear as presented in Fig. 3 b) and c). Then, all the wear was detected on the HS25 sample (missing volume) and wear volume was defined so that:

$$V = V_{HS25}^- \qquad (4)$$

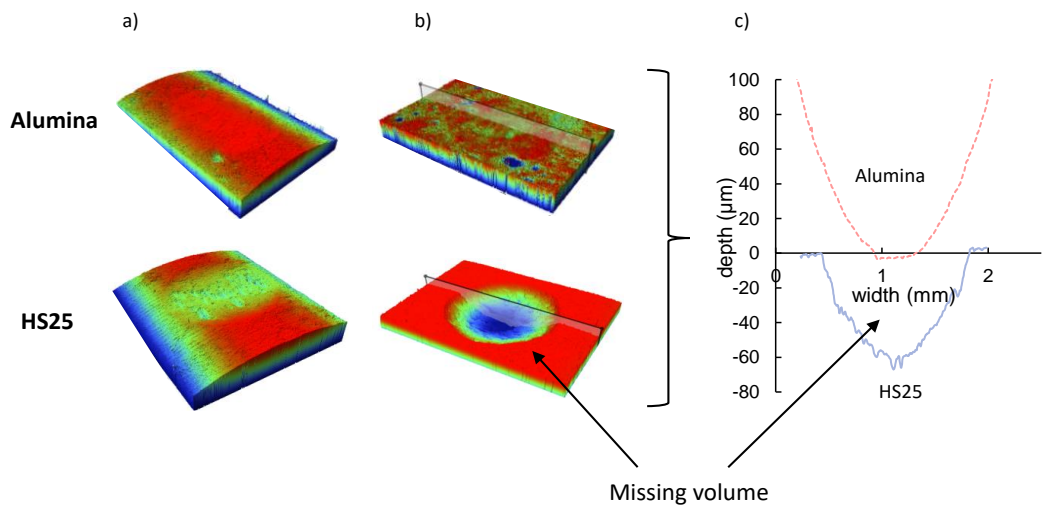

Fig. 3: a) 3D profilometry of the fretted interface; b) After cylinder shape removing; c) Equivalent 2D profiles (reprinted from [18] with permission from Elsevier)

*1.4.2. Microstructure and chemical observations*

The wear tracks were observed with an optical microscope and a SEM (FEG Mira 3 TESCAN) using an accelerating voltage of 5 kV. The EDX quantifications were performed with an Oxford Instruments system at 20 kV. Cross-sections were made mechanically (cutting and grinding) or by FIB machining with a Dual Beam FEI Helios 600i. A TEM lamella was also observed and the description of the method is available in [21].

*1.4.3. Micro-mechanical tests*

The high-temperature tribolayer and HS25 mechanical properties were determined using the micro-pillar compression technique as already used by some authors [15,23,24]. The device used was an Alemnis high-temperature *in situ* indenter [25] installed in a Tescan VEGA SEM. Micro-pillar compression was performed using a 5 µm heated flat tip after a temperature calibration procedure of both the tip and the sample. The methodology used is the same as in [15] where pillars were machined using a gallium ion source on a cross-sectioned and polished sample. The pillars had a diameter of 1 µm and a height of 2 µm. Pillars were compressed at room temperature, 100°C, 200°C, 300°C and 400°C at a constant strain rate (0.001 $s^{-1}$). The engineering stress was calculated by dividing the measured load by the pillar upper surface area and the strain by dividing the displacement by the pillar initial height.

2. Tribological results

   2.1. Effect of temperature on wear

As previously observed on this tribosystem [18,19,21], the temperature has a great effect on the wear mechanism. Fig. 4 a) shows the evolution of wear volume and the friction coefficient as a function of the operated temperature, while Fig. 4 b) and c) respectively display the related wear kinetics and the wear scar associated to each domain (reference temperatures). Three domains can be distinguished:

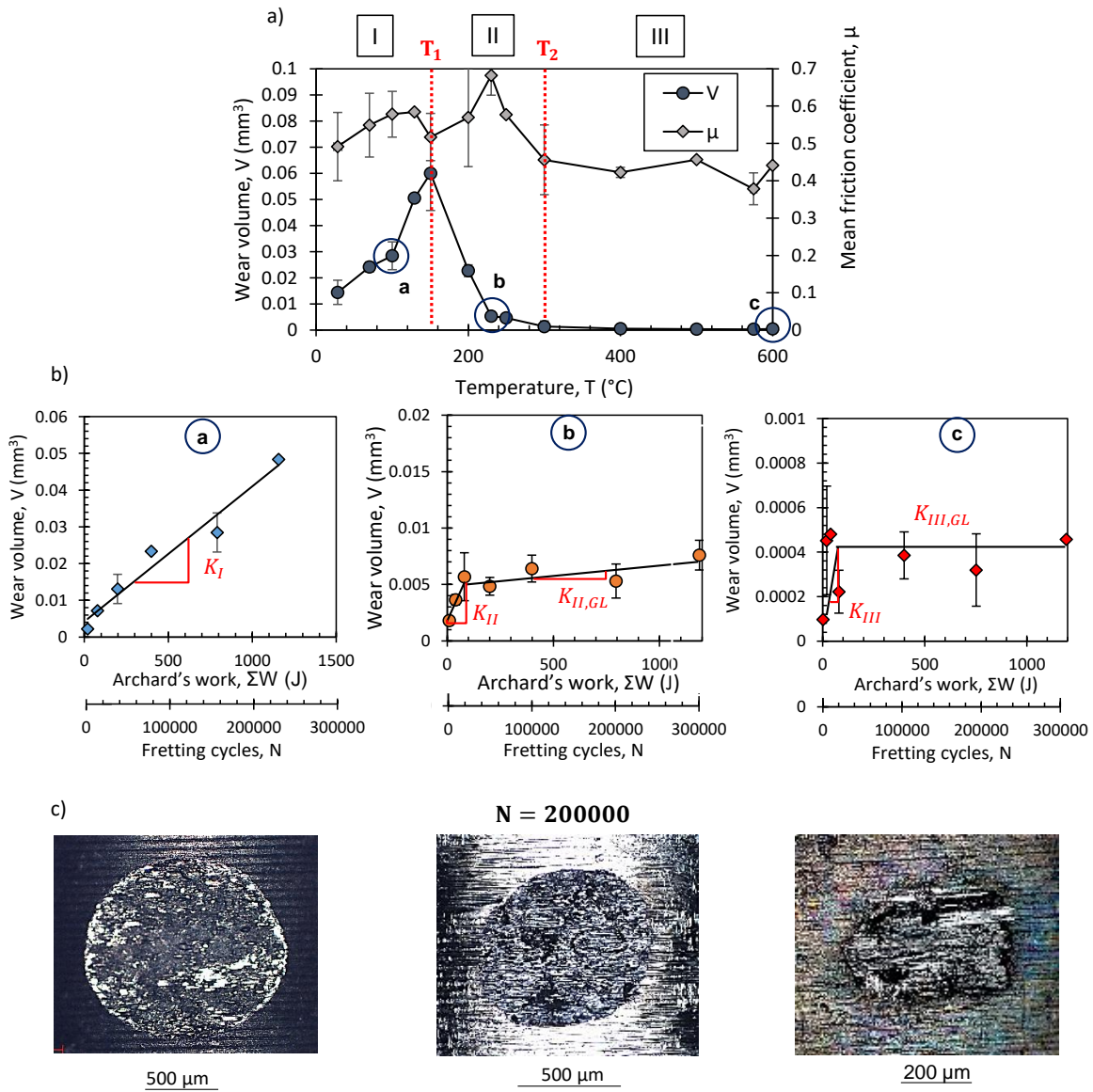

Fig. 4: a) Effect of the temperature on the wear volume and the friction coefficient (f = 50 Hz, P = 50 N, N = 200 000 cycles, $\delta_0$ = ± 20 µm) (partially reprinted from [21] with permission from Elsevier); b) Wear kinetics for the three reference temperatures : 100°C, 230°C and 575°C ($K_I$ = $3.68.10^{-5}$ mm$^3$/J, $K_{II}$ = $1.28.10^{-5}$ mm$^3$/J, $K_{II,GL}$ = $2.06.10^{-6}$ mm$^3$/J, $K_{III}$ = $5.14.10^{-7}$ mm$^3$/J, $K_{III,GL}$ = $2.38.10^{-8}$ mm$^3$/J) ; c) Optical images of the associated wear tracks for the reference temperatures (f = 50 Hz, P = 50 N, $\delta_0$ = ± 20 µm, N = 200 000)

- Domain I (T ≤ $T_1$): this domain is characterized by a continuous increase in the wear volume up to a maximum value achieved for T = 150°C. The friction coefficient is quite high and constant. The wear kinetics (Fig. 4 b)) drawn when T = 100°C shows a linear increase in the wear volume between 0 and 300 000 cycles. The Archard's wear coefficient for domain I, $K_I$, is constant and equals to $3.68.10^{-5}$ mm$^3$/J. Finally, the wear scar presented in Fig. 4 c) exhibits abrasive grooves with the absence of any tribolayer;

- Domain II ($T_1$ < T < $T_2$): this domain displays a sharp decrease in the wear volume until a very low value at 300°C. The wear kinetics established at 230°C displays a bi-linear evolution. First, the wear volume increases linearly with fretting cycles ($K_{II}$ = $1.28.10^{-5}$ mm$^3$/J) until a critical number of fretting cycles is achieved. Then, the wear volume remains almost constant and the slope in Fig. 4 b) is almost equal to zero such as $K_{II,GL}$ = 2.06.10$^{-6}$ mm$^3$/J. This means that the production of wear debris is drastically reduced at the interface. The associated wear track reveals the presence of a bright layer at the interface, protecting it from severe wear. The friction coefficient is also decreasing, after showing a slight increase while wear volume is decreasing. This phenomenon is still unclear but is discussed elsewhere [19].

- Domain III (T ≥ T$_2$): this domain is characterized by a very low wear volume. The wear kinetics at 575°C shows the same bi-linear evolution as in domain II but with no wear variation at all when the critical number of fretting cycles is achieved ($K_{III}$ = 5.14.10$^{-7}$ mm$^3$/J, $K_{III,GL}$ = 2.38.10$^{-8}$ mm$^3$/J). The associated wear track is also covered by a bright layer usually called "glaze layer" leading to the complete protection of the fretted interface [7].

2.2. Glaze layer formation (domain III, $T > T_2 = 300°C$)

Fig. 5 a) presents the evolutions of the friction coefficient and the wear volume at 575°C (domain III) with the fretting cycles. At such a high temperature, it was shown that a protective glaze layer is formed leading to a drastic reduction of wear [19].

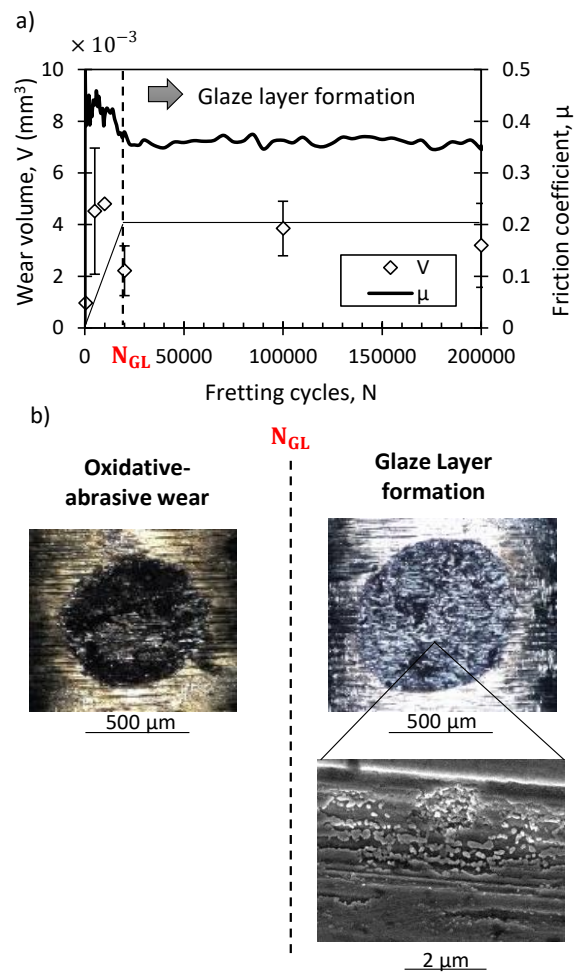

Fig. 5: a) Evolutions of the wear volume and the friction coefficient with fretting cycles at 575°C (III); b) Optical images of the wear tracks before and after the glaze layer formation and SEM images of the sintering process resulting in glaze layer (f = 50 Hz, P = 50 N, δ$_0$ = ± 20 µm, T = 575°C)

In the beginning of the high-temperature fretting test, the coefficient of friction increases up to a maximum value (0.45) and then starts to decrease until a low constant value is achieved (0.36) after 20 000 cycles. In the same time, the wear volume increases (for N < 20 000) and then remains constant (for N ≥ 20 000). A critical number of fretting cycles $N_{GL}$ can be considered to represent the number of fretting cycles after which the coefficient of friction is stabilized and the wear volume remains constant (i.e when the wear rate tends to zero). According to Fig. 5 a), at 575°C, it can be stated that the protective glaze layer is effective after about 20 000 cycles since the interface does not present further damage afterwards. This is corroborated by Fig. 5 b) showing that the interface is covered by a powdered debris-bed layer when $N < N_{GL}$ and by a bright glaze layer when $N \geq N_{GL}$. The SEM image presented in Fig. 5 b) is a detail of the glaze layer interface and shows that it is formed by the compaction and the sintering of wear debris particles [10].

### 2.3. Structure and chemical composition of the glaze layer

#### 2.3.1. Microscopic observations

Cross-sections and associated SEM/EDX analyses were performed in the wear track and are presented in Fig. 6 a) before and after glaze layer formation. Fig. 6 b) presents a schematization of the interfaces as well as the evolution of the [Co]/[Cr] ratio measured by EDX (at. %). The [Co]/[Cr] ratio is used to interpret the evolution of the concentration of the main chemical elements of the alloy. Initially, [Co]/[Cr] is equal to 2 in the HS25 alloy. When [Co]/[Cr] > 2, it means that there is a cobalt enrichment compared to the initial ratio and when [Co]/[Cr] < 2, there is a chromium enrichment.

Before glaze layer formation, the cross-section reveals that the interface is oxidized through a depth of 2-3 µm and presents some porosities. The [Co]/[Cr] ratio shows that there is a chromium enrichment at the oxide/bulk interface referred to as "Chromium-Rich Layer" or "CRL". Above the CRL, the interface is oxidized with a gradient of cobalt enrichment from the bulk material towards the surface. This layer is called "Mixed Oxide Layer" (MOL) [21]. Below the CRL stands the bulk material, namely the HS25, where the [Co]/[Cr] ratio remains constant and equal to 2. The wear track after the glaze layer formation is quite similar displaying also the MOL and the CRL structures. The MOL displays a [Co]/[Cr] ratio equals to the initial one (2) which means that there is the same concentration in cobalt and chromium as in the original HS25 material. However, a new layer, less than 1 µm thick, is observed at the top of the surface. The top layer is enriched in cobalt and is therefore called "Cobalt-Rich Layer" (CoRL). High-resolution XPS spectra were previously performed [21] inside the wear tracks from top view before ($N < N_{GL}$) and after ($N \geq N_{GL}$) the glaze layer formation for cobalt $2P_{3/2}$ and chromium $2P_{3/2}$. The main peaks of both spectra were shifted toward the highest values of binding energy which means that the elements are principally oxidized. From this data, it was concluded that there is a cobalt enrichment when $N \geq N_{GL}$ and a chromium enrichment when $N < N_{GL}$ (Fig. 6 b)).

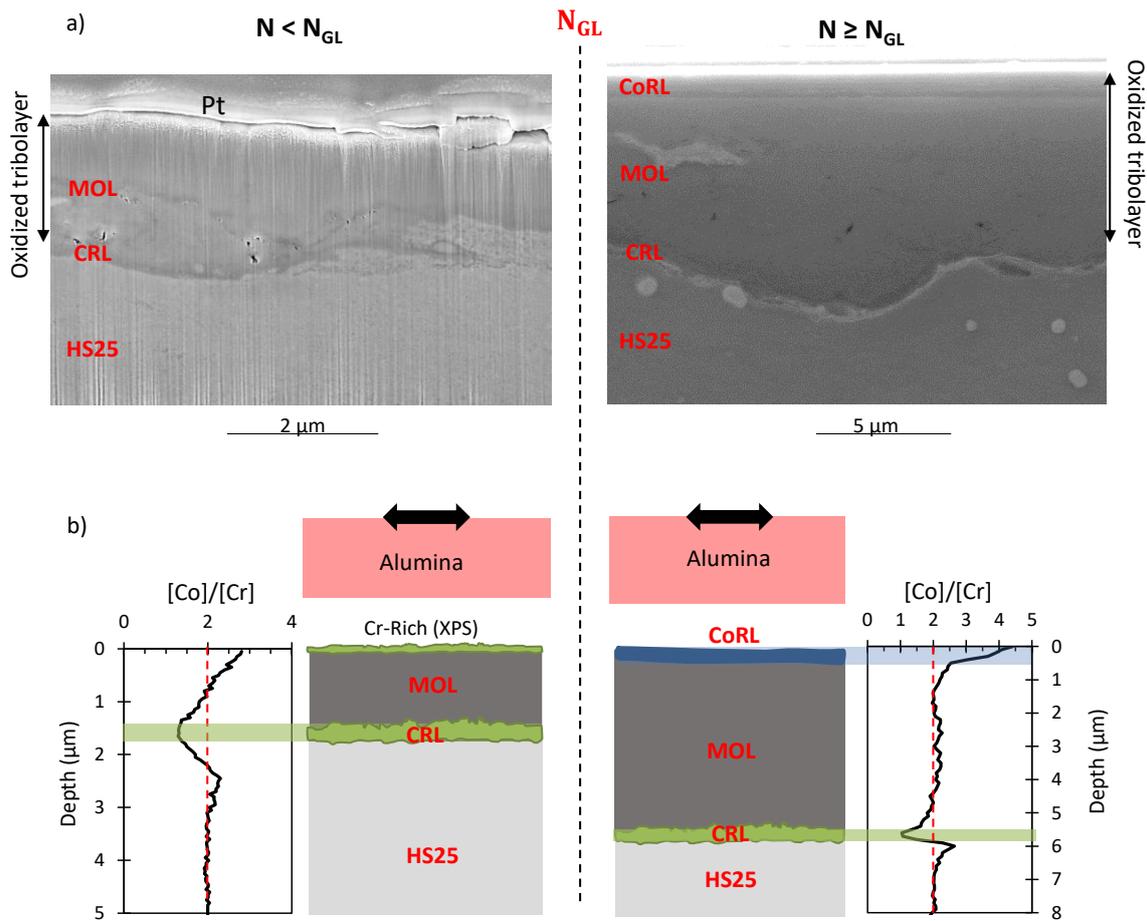

Fig. 6: a) Cross-section SEM images of the fretted interface before (N = 5 000 < $N_{GL}$) and after (N =200 000 ≥ $N_{GL}$) the glaze layer formation (f = 50 Hz, P = 50 N, $\delta_0$ = ± 20 µm, T = 575°C); b) Schematization of the interfaces and evolution of the [Co]/[Cr] ratio

### 2.3.2. High resolution observations

A TEM lamella was extracted from the middle of a typical wear track covered by the protective glaze layer (N = 200 000, T = 575°C) as presented in Fig. 7 a). The CoRL and the MOL are clearly visible on the lamella and display different textures. The CoRL has a thickness of about 600 nm and is quite homogeneous. Fig. 7 b) shows details of the CoRL where the grains are nanocristalline and display a constant grain size around 40-50 nm [21]. Moreover, the juxtaposition of the grains indicates a sintered structure. Electronic diffraction was previously performed in the CoRL ([21] and Fig. 7 b)) and showed that there is no preferential orientation of the grains and probably a very limited number of chemical species are present since the rings are made of isolated spots. According to the radii of the rings, the CoRL is made of $Co_3O_4$ cobalt oxides [21]. Finally, the MOL in Fig. 7 a) is more heterogeneous with a mixture of different grain sizes and some porosities (white color).

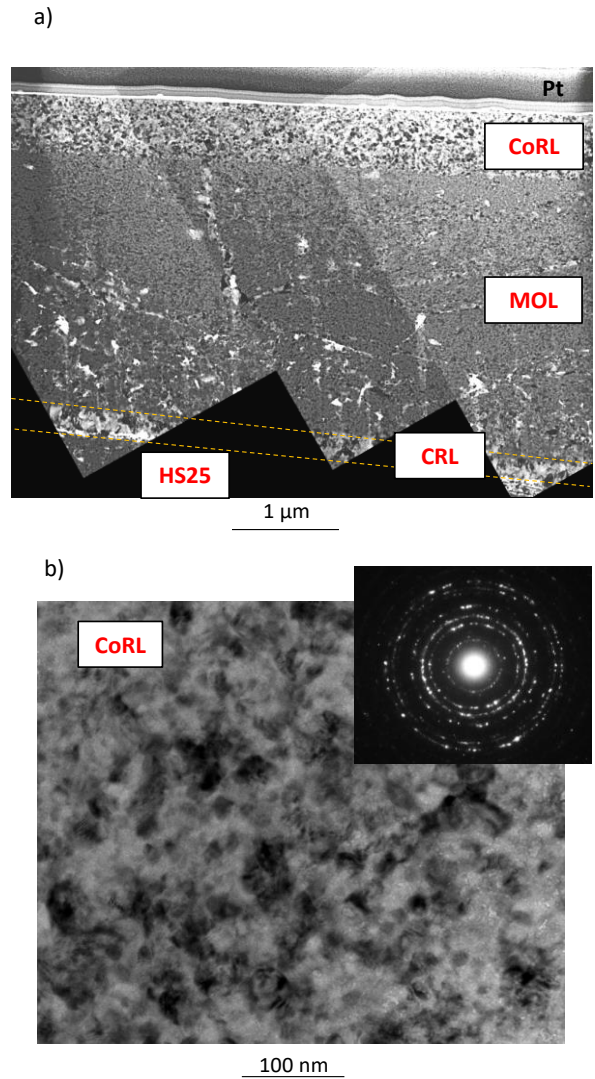

Fig. 7: a) TEM image of the lamella in bright field (reprinted from [21] with permission from Elsevier); b) Bright field observation of the CoRL and diffraction pattern (f = 50 Hz, P = 50 N, N = 200 000 cycles, $\delta_0$ = ± 20 µm, T = 575°C)

### 2.3.3. Wear kinetics

In the light of these results, the importance of each layer in the protective wear process is discussed. Since the CoRL is the major difference between the fretting interface when $N < N_{GL}$ and $N \geq N_{GL}$, it is highly probable that the CoRL is the layer that is able to protect the interface from wear, namely the glaze layer. The sintered aspect of the CoRL (Fig. 7 b)) confirms that this layer is formed by a tribo-sintering process of the debris particles in the interface.

Experiments indicated that the chemical composition of the surface is changing from a cobalt depletion when $N < N_{GL}$ to a chromium depletion when $N \geq N_{GL}$. Since the effective glaze layer (CoRL) is found to be richer in cobalt than the bulk material, a selective process encouraging the tribo-sintering of cobalt oxides instead of chromium oxides may occur (Fig. 6 b)). Based on these considerations and a previous study [21], a wear scenario at high temperature is proposed in Fig. 8:

a. In the beginning of the test, when N < $N_{GL}$, the interface is richer in chromium than in cobalt. According to a previous study [21], this may be due to the higher oxidation rate of chromium compared to cobalt which then enhanced the transient oxidation of chromium instead of cobalt. Hence, the interface is enriched in Cr elements which are continuously worn out and ejected;
b. The depletion in chromium leads to the production of a bigger quantity of cobalt oxides which are also worn out. Cobalt oxides are not ejected out of the interface but rather compacted due to a better diffusional behavior of this oxide compared to Cr oxide [11,21]. This selective diffusion process is not well formalized for now and will be discussed in section 3;
c. Finally, the glaze layer is formed by a tribo-sintering process and it is mainly constituted of cobalt oxides. The effectiveness of the glaze layer will also be discussed in section 3.

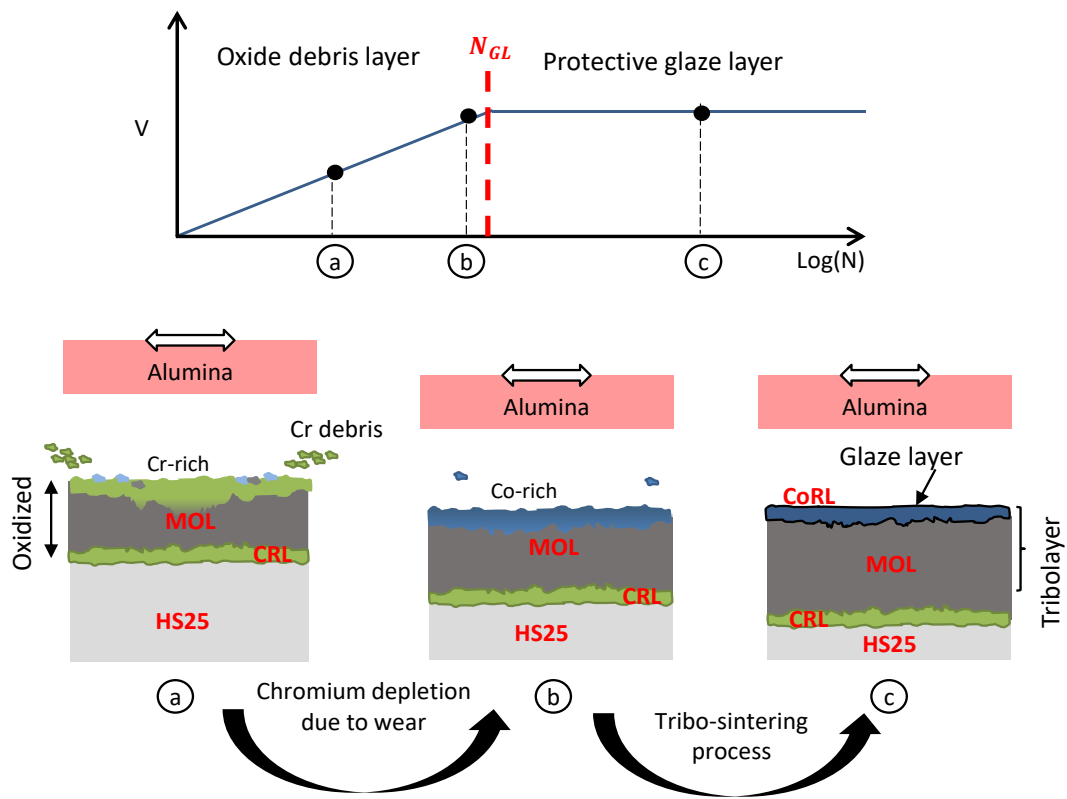

Fig. 8: Evolution of the interface subjected to fretting-wear at high temperature (reprinted and partially modified from [21] with permission from Elsevier)

The structure of the high temperature tribolayer and its evolution towards an effective glaze layer with the formation of a thin protective CoRL structure is now described. To complete the description of the oxidized tribolayer, the mechanical properties of the fretted interface are presented.

2.4. Mechanical properties of the fretted interface

2.4.1. Micro-pillars compression

Many authors argue that the exceptional tribological properties of the glaze layer is due to its mechanical properties. Some authors [14,15] showed that the glaze layer is very hard compared to the bulk material which may promote a reduction of the wear rate.

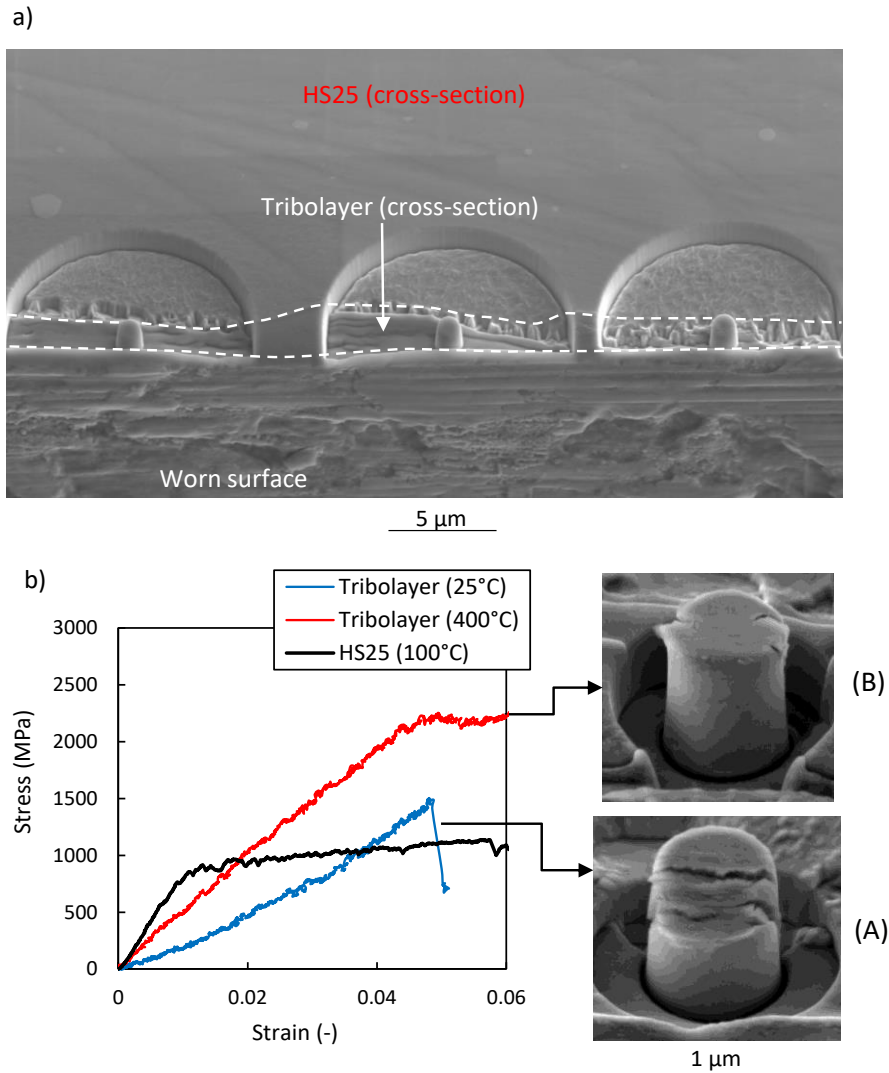

Fig. 9: a) SEM images of three pillars machined in the fretted interface; b) Stress-strain curves for micro-compression of pillars machined in the Mixed Oxide Layer (MOL) and in the bulk material (HS25)

More recently, Viat *et al.* [15] machined micro-pillars in the tribolayer and compressed them at different temperatures. The results showed a fragile/ductile transition and the authors proposed that the absence of fragility of the tribolayer at high temperature drastically limits the debris production and then the debris ejection flow tends to zero. In the present paper, it was showed that the high temperature tribolayer is not homogeneous, but rather composed of three layers (CRL, MOL and CoRL, as observed in Fig. 7). Even if it is believed that the glaze layer, i.e. the layer able to protect the interface from further wear, is only the very thin Cobalt-Rich Layer, it could be interesting to obtain the mechanical properties of the thick oxidized layer. Following the same method as in [15], some micro-pillars were machined in the tribolayer formed at high temperature (mainly composed of the MOL) as showed in Fig. 9 a) and compressed at different temperatures, from ambient to 400°C. Due to the experimental constraints, the compression of pillars at higher temperatures was not possible. However, Fig. 4 a) shows that the wear volume and the friction coefficient are already low at 400°C and it is assumed that the wear behavior of the fretted interface at 400°C is equivalent to that of the fretted interface at 575°C. Fig. 9 b) presents two typical stress-strain curves for the tribolayer (100°C and 400°C) and a typical curve for the bulk material (100°C). The typical curve obtained for the HS25 bulk material is also provided for comparison.

For the oxidized tribolayer, it is observed that at low temperature (100°C) the curve presents a sudden drop in the stress-strain curve which indicates some crack propagation through the pillars [15]. SEM observations confirmed the presence of longitudinal cracks when the pillar is tested at low temperature and thus confirming the fragile behavior but also a layered structure of the tribolayer (A). No sudden drop occurs at high temperature, showing a ductile-like behavior. SEM observations also confirmed the ductility of the pillar showing a barrel-like deformation (B). The Haynes 25 displays an elastic-plastic strain-stress curve at 100°C as previously detailed in [15,26]. This confirms that the Haynes 25 changes from a ductile behavior to a fragile behavior due to the tribological transformation. Finally, it is observed that the yield strength, and therefore the hardness, of the oxidized tribolayer are higher than those of the bulk material.

### 2.4.2. Discussion

In [15], authors made a correlation between the ductility of the tribomaterial at high temperature and the capacity of the glaze layer to resist fretting wear. They proposed to quantify the brittleness of the pillars by using statistical methods (pillars with brittleness in %). Indeed, they showed that for the same temperature, the behavior may be different (ductile or fragile) from one pillar to another. However, at 500°C and above, all the pillars displayed a ductile behavior which hence may explain the resistance of the glaze layer to wear. They assumed that the capacity of the glaze layer to resist fretting-wear is related to the ductility of the layer allowing a plastic accommodation of the fretting cyclic strain.

Fig. 10 compares the evolution of the wear volume (data from Fig. 4) and the evolution of the statistical brittleness behavior of the pillars as a function of the temperature. In this study, the statistics are made on a very small number of data (2 or 3) due to the difficulty of performing such experiments. It is observed that all the pillars exhibit a ductile behavior at 400°C which is also the temperature at which the glaze layer is formed so that the wear rate converges to zero ($K_{III,GL} = 0$). These results are in accordance with those of Viat *et al.* [15], and seems to confirm a link between the protective role of the glaze layer and its ductility. However, the glaze layer is not homogeneous as previously depicted in Fig. 6 b) and Fig. 8 b) and the pillars were preferentially machined in the Mixed Oxide Layer (MOL). Microstructural observations at different number of fretting cycles showed that the effective glaze layer, formed when N ≥ $N_{GL}$, consists of a very thin Cobalt-Rich Layer (CoRL) on the top surface. Hence, the tribological role of the MOL is uncertain but its mechanical behavior may be close to that of the CoRL layer since both are oxidized and nanocrystalline. Further efforts need to be made in order to have a better description of the mechanical behavior of the CoRL, by performing for example some *in situ* scratch experiments at high temperature.

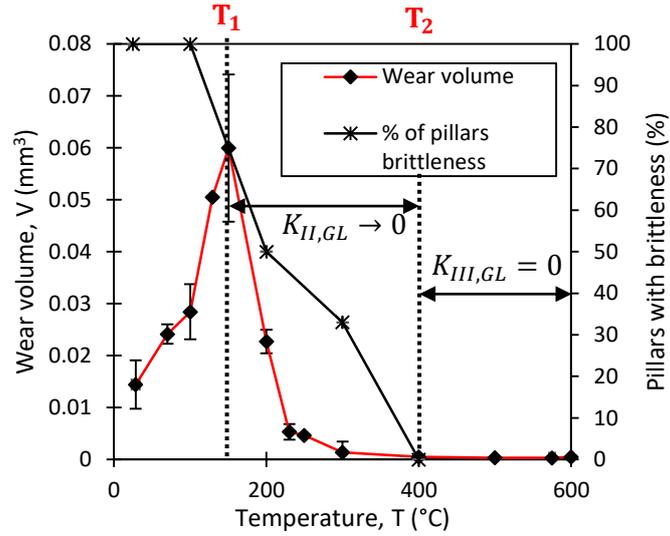

Fig. 10: Correlation between the glaze layer formation (T ≥ T$_2$) and the mechanical behavior of the entire tribolayer represented by the percentage of pillars exhibiting a brittle behavior (as proposed by Viat et al. in [15])

In this section, the oxidized tribolayer obtained at high temperature was morphologically, chemically and mechanically characterized. In the next section, the focus is set on the sintering process that involved in the formation of the glaze layer.

## 3. Formalization of the tribo-sintering process

It is now proposed to formalize the glaze layer formation through a tribo-sintering concept as previously performed by Kato and Komai [10]. According to the authors, the wear volume at high temperature is directly dependent on the tribolayer formation (since there is no additional wear when the glaze layer is effective) and is inversely proportional to the sintering rate $S$ of the triboparticles:

$$V \propto L_{GL} \propto \frac{1}{S} \qquad (5)$$

Where $V$ is the wear volume and $L_{GL}$ the sliding distance until the glaze layer is formed.

### 3.1. Formulation of N$_{GL}$

The investigation of the tribo-sintering process is performed through the analysis of the $N_{GL}$ variations, which are directly linked to the ability of the tribosystem to form the glaze layer. In the following, the $N_{GL}$ values were determined by considering that the stabilization of the friction coefficient towards its lower value is the consequence of the glaze layer formation and then indicates the number of fretting cycles at which the glaze layer is formed, $N_{GL}$ (Fig. 5 a)). Fig. 11 shows the N$_{GL}$ variations for different tribological parameters ($T, f, \delta_0$).

#### 3.1.1. Influence of the frequency

Fig. 11 a) plots the influence of the frequency on $N_{GL}$ shows a linear evolution. The frequency is directly related to the time during which the triboparticles undergo sintering. By assuming that the system needs a minimal time $t_{GL}$ to achieve the tribo-sintering process and form the glaze layer, the following equation is proposed:

$$t_{GL} = \frac{N_{GL}}{f} \quad (6)$$

It is also considered that *S*, the sintering rate introduced by Kato *et al.* is able to quantify the tribo-sintering process during the experiment so that when $N = N_{GL}$, *S* is associated to a given value noted $S_{GL}$.

As classically observed for the sintering of particles, it is assumed that the sintering rate is directly proportional to the process time:

$$S \propto t \propto \frac{N}{f} \quad (7)$$

And then,

$$S_{GL} \propto t_{GL} \propto \frac{N_{GL}}{f} \quad (8)$$

### 3.1.2. Influence of the temperature

Fig. 11 b) shows that the temperature has a strong influence on $N_{GL}$. The higher the temperature, the faster the formation of the glaze layer. An asymptotic evolution of $N_{GL}$ is observed regarding the temperature which is consistent with a sintering approach [10]. The following relation is proposed:

$$S_{GL} \propto \exp\left(-\frac{E_{a,GL}}{RT}\right) \times \frac{N_{GL}}{f} \quad (9)$$

Finally, we get the following expression of $N_{GL}$:

$$N_{GL} \propto \frac{S_{GL} \times f}{\exp\left(-\frac{E_{a,GL}}{RT}\right)} \quad (10)$$

### 3.1.1. Influence of the sliding amplitude

Finally, Fig. 11 d) shows the influence of the sliding amplitude on $N_{GL}$. A decrease in $N_{GL}$ is observed when the sliding amplitude increases. According to Kato *et al.* [10], the sliding distance leading to the glaze layer $L_{GL}$ is an important parameter to take into account in the description of the glaze layer formation through a tribo-sintering process (Eq. (5)). $L_{GL}$ can be formulated as follows:

$$L_{GL} = 4\delta_0 N_{GL} \quad (11)$$

Eq. (11) implies that:

$$N_{GL} \propto \frac{1}{\delta_0} \quad (12)$$

Experimentally, it is observed that $N_{GL}$ depends on the inverse of the square root of the sliding amplitude (Fig. 11 d)):

$$(N_{GL})_{exp} \propto \frac{1}{\sqrt{\delta_0}} \qquad (13)$$

This dependency is not fully understood for now and additional investigations are needed to confirm this particular point. In addition, the global wear model proposed in section 4 showed that Eq. (13) leads to a better description of the wear behavior than Eq. (12).

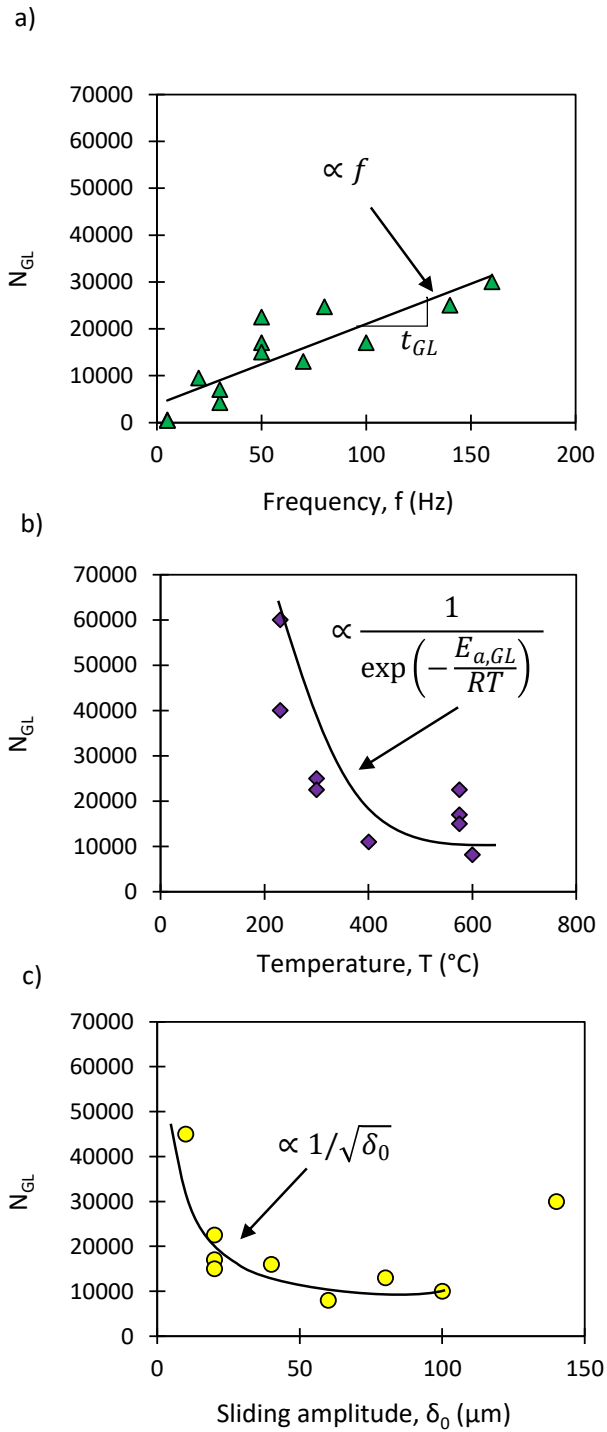

Fig. 11: Influence of various tribological parameters on the $N_{GL}$ number: a) frequency; b) temperature; c) sliding amplitude ($f_{ref}$ = 50 Hz, $P_{ref}$ = 50 N, $\delta_{0,ref}$ = ± 20 µm, $T_{ref}$ = 575°C, N = 200 000)

### 3.1.2. Experimental fitting

In the light of these considerations, a formulation of $N_{GL}$ is proposed in Eq. (14). Note that this formulation is based on a physical description of the sintering process of the triboparticles except for the sliding amplitude whose influence is captured experimentally.

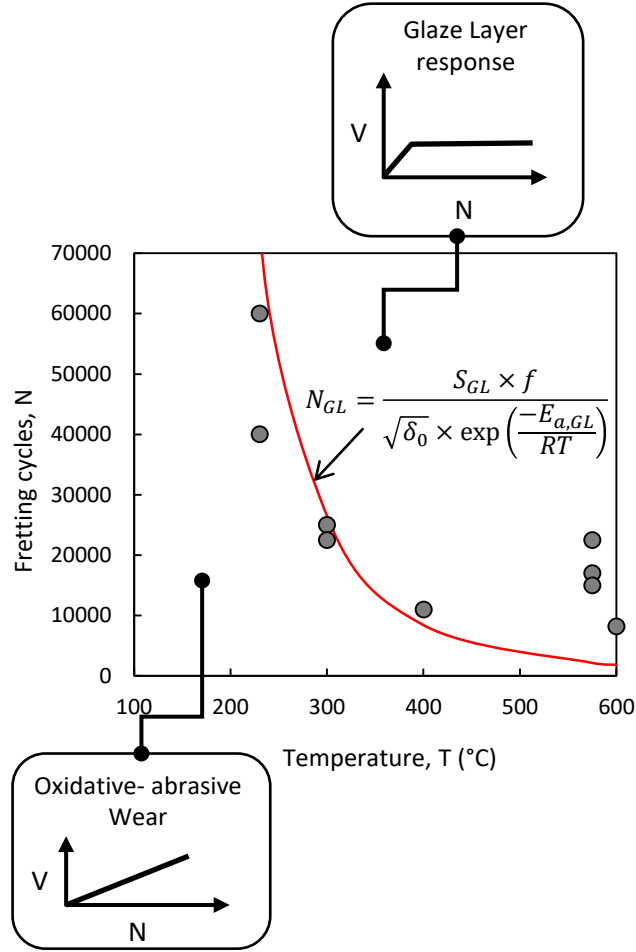

Fig. 12: Evolution of the $N_{GL}$ number with the operating temperature (f = 50 Hz, P = 50 N, $\delta_0$ = ± 20 µm, N = 200 000)

$$N_{GL} = \frac{S_{GL} \times f}{\sqrt{\delta_0} \times exp\left(-\frac{E_{a,GL}}{RT}\right)} \qquad (14)$$

Where $S_{GL}$ is a fitting constant (s.mm$^{0.5}$). Eq. (14) is adjusted by considering the evolution of $N_{GL}$ with the temperature. Fig. 12 shows the results with $S_{GL}$ = 0.015 s.mm$^{0.5}$ and $E_{a,GL}$ = 40000 kJ/mol . A good correlation is observed between the theoretical $N_{GL}$ values and the experimental values, except for the data at high temperature where the number of fretting cycles $N_{GL}$ is underestimated. However, the wear model developed in section 4 shows that the best correlation is observed for the present values of $S_{GL}$ and $E_{a,GL}$. Finally, Fig. 12 shows that, using such a formulation, it is possible to distinguish the severe wear domain I (oxidative-abrasive) from the mild wear domains II and III (formation of a protective tribolayer). Hence, $N_{GL}$ appears to be a good indicator to know how the contact behaves under fretting wear.

## 3.2. Tribo-sintering process

A formulation of $N_{GL}$ is established, based on a sintering process of the triboparticles, and also supported by experimental considerations. It is now proposed to strengthen such hypotheses by considering the description of the sintering process of powders.

### 3.2.1. Sintering rate

Sintering is a process where particles are welded to each other by diffusion. One of the main steps of the sintering process is the formation and the growth of bridges among the contacting particles. The diffusion mechanism involved in the sintering process is dependent on the operating temperature and can take place at the surface, at the grain boundaries or in the lattice paths [20]. The speed of the bridges' building can be measured by considering the parameter $\lambda$ standing for the evolution of the interpenetration of two perfect spheres (Fig. 13).

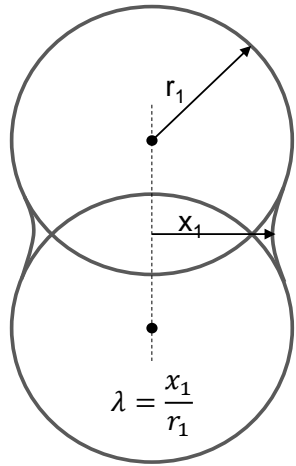

Fig. 13: Interpenetration of two perfectly spherical particles during the sintering process

Hence, a way to calculate the sintering rate is to follow the evolution of $\lambda$ with time, expressed by the following equation [27]:

$$\lambda^n = \frac{k_i D_i}{r^m} t \qquad (15)$$

Where $k_i$ is a constant depending on the involved diffusion mechanism $i$, $D_i$ is the diffusion coefficient of the mechanism $i$, $r$ is the sphere radius, $m$ and $n$ are constants, and $t$ the time allowed for the process to take place. By introducing the temperature dependence of the diffusion process Eq. (15) becomes:

$$\lambda^n = \frac{k_i D_{0,i} \exp\left(-\frac{E_{f,i}}{RT}\right)}{r^m} t \qquad (16)$$

With $D_{0,i}$ a pre-exponential factor and $E_{f,i}$ the activation energy of the involved diffusion mechanism $i$.

### 3.2.2. Application to the tribo-sintering process

Obviously, there is some similitude between Eq. (9) and Eq. (16). If we consider that the time $t$ in Eq. (16) is the time necessary to sinter the cobalt oxide debris particles $t_{GL}$, defined in Eq. (6), we obtain the following formulation:

$$\lambda_{crit}^n \propto \frac{k_i D_{0,i} \exp\left(-\frac{E_{f,i}}{RT}\right) N_{GL}}{r^m f} \quad (17)$$

Where $\lambda_{crit}^n$ can be defined as the minimum inter-penetration required between two particles to withstand the sliding solicitation. Moreover, Eq. (17) implies that we consider that the inter-penetration of two particles is representative of the global sintering process of the glaze layer and that all the particles sinter at the same time $t_{GL}$. This hypothesis is corroborated by the fact that experimentally, the glaze layer seems to be created quasi-instantaneously at $t_{GL} = N_{GL}/f$ (Fig. 5). Finally, from Eq. (17) we get:

$$N_{GL} \propto \frac{\lambda_{crit}^n f r^m}{k_i D_{0,i} exp\left(-\frac{E_{f,i}}{RT}\right)} \quad (18)$$

This formulation of $N_{GL}$ is close to the one obtained by the parametric analysis of Eq. (14) except that there is no dependence with the sliding amplitude in Eq. (18) and that the dependence on the size of the wear debris particles is taken into account. $N_{GL}$ is now, with this formulation, related to the diffusion properties of the oxides (diffusion coefficient, $D_i$). This formulation strengthen the close relation between the diffusive properties of the oxides present at the interface and the formation of the glaze layer. Eq. (18) shows that the higher the diffusion coefficient of the oxides and the faster the formation of the glaze layer by lowering the $N_{GL}$ parameter.

### 3.3. Role of the alloying elements

To confirm the selective sintering process involved in the formation of the cobalt-based glaze layer, as intuited in section 2, the role of chromium and cobalt oxides in the glaze layer formation was analyzed by performing fretting-wear on pure cobalt and pure chromium [11]. The aim of this study was to analyze the high temperature tribological response of three different tribosystems involving pure cobalt and pure chromium rods (Fig. 14 a)). Mechanical properties of the pure materials are presented in [11] and the tribological parameters were the same as those of the HS25//alumina tribosystem (f = 50 Hz, P = 50 N, $\delta_0$ = ± 20 µm, T = 550°C, N = 200 000 cycles).

#### 3.3.1. Tribological results

Fig. 14 presents the tribological results for the three tested tribocouples (Co//Cr, Co//Co, Cr//Cr) in terms of friction coefficient (Fig. 14 b)) and fretting scar observations (Fig. 14 c)). It is noticed that a glaze layer is formed in the Co//Co and Co//Cr tribosystems, since a bright layer is optically observed in Fig. 14 c) for these two tribocouples but not for the Cr//Cr one. This is also confirmed by the analysis of the friction coefficient. Indeed, the friction coefficient is high and not stable for Cr//Cr whereas a drop and a stabilization of µ is observed for Co//Co and Cr//Co. The instability of the friction coefficient for the Cr//Cr tribocouple probably show the incapacity of the interface to create a stable protective third body, which is rather continuously broken into debris and ejected out of the interface.

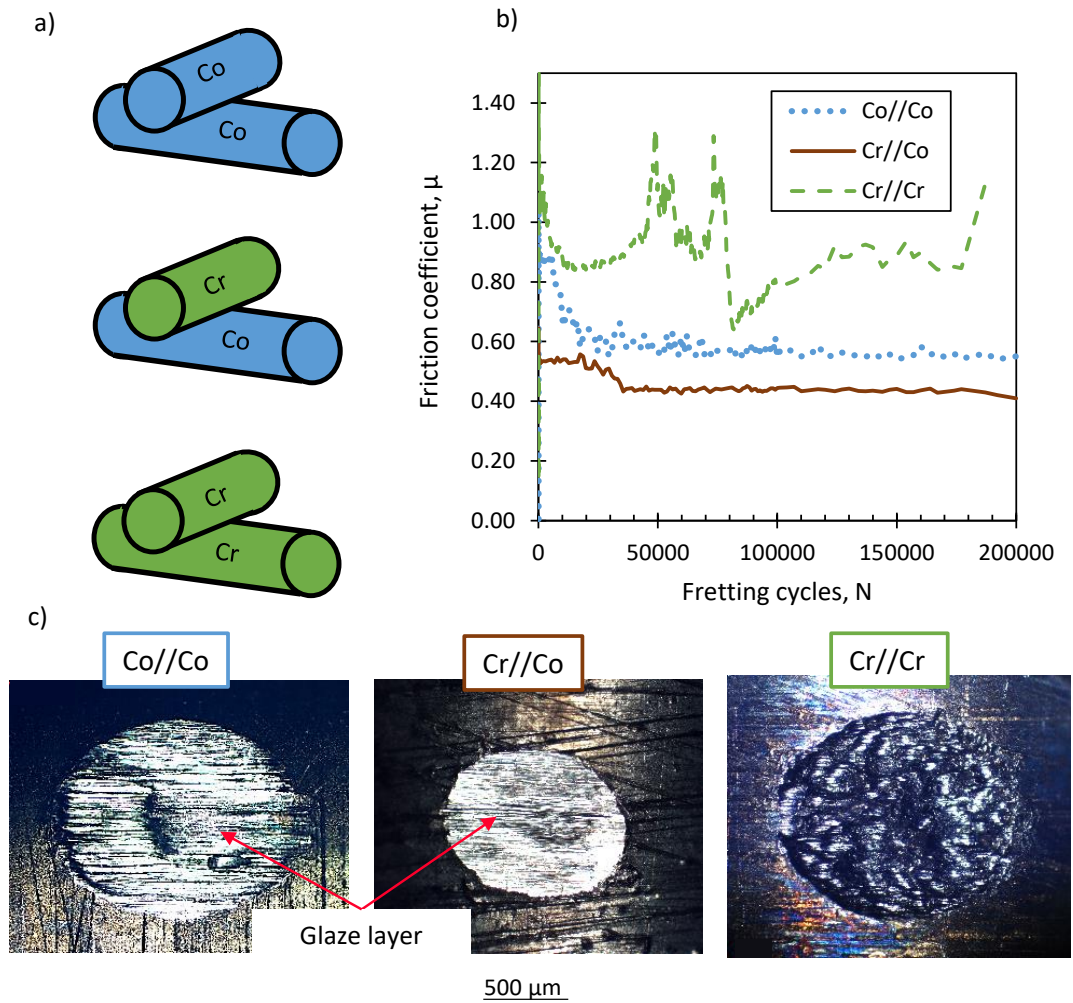

Fig. 14: a) Illustration of the tribocouples involving pure metallic elements tested under fretting-wear at high temperature; b) Evolution of the friction coefficients for the three studied tribocouples (Co//Co; Cr//Cr, Co//Cr) (reprinted from [11] with permission from Elsevier); c) Optical images of the wear tracks (f = 50 Hz, P = 50 N, $\delta_0$ = ± 20 µm, T = 550°C, N = 200 000 cycles)

Fig. 15 shows some Raman analyses performed in the wear tracks [11]. It is observed that the Raman spectra for the Co//Co and Co//Cr tribocouples are similar and Raman peaks are those of $Co_3O_4$. The Raman spectrum performed in the Cr//Cr wear track has a different shape and appears to be close to that of $Cr_2O_3$. Hence, the glaze layer formed on Co//Cr and Co//Co seems to be constituted of the same cobalt oxides ($Co_3O_4$), as those of the HS25//alumina tribocouple (TEM characterizations presented in Fig. 7).

Thus, cobalt plays a key role in the formation of the glaze layer since it seems to be the necessary chemical element to form the glaze layer, among all the chemical elements present at the interface. Moreover, according to Eq. (18), $N_{GL}$ is linked to the diffusion properties of the particles. Hence, a physical relation can be made between the diffusion properties of the chemical elements and their capacity to form a glaze layer. The role of the alloying elements regarding the glaze layer formation is then discussed.

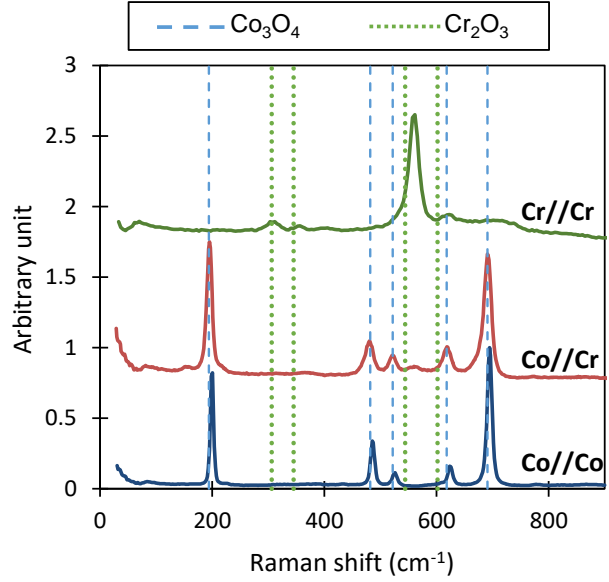

Fig. 15: Raman spectra performed on the high temperature wear tracks for Co//Co, Cr//Cr and Co//Cr tribocouples (reprinted from [11] with permission from Elsevier)

### 3.3.2. Discussion on the role of the alloying elements

Concerning the Co//Co, Co/Cr and Cr//Cr tribosystems, sintering could happen between two cobalt oxide particles, two chromium oxide particles or the combination of both oxides. Oxygen ions and metallic ions can diffuse from one grain to another. Since metallic ions are smaller than oxygen ions, it is more likely that cationic diffusion is the main diffusion phenomenon involved in the tribo-sintering process. Moreover, it is also proposed that the diffusion process presenting the highest diffusion coefficient is more likely to be dominant and form the glaze layer than others so that:

$$D_i = \max(D_{Co \rightarrow Co_3O_4}; D_{Cr \rightarrow Cr_2O_3}; D_{Cr \rightarrow Co_3O_4}; D_{Co \rightarrow Cr_2O_3}) \quad (19)$$

Table 3 presents some data found in the literature about the auto-diffusion of metallic ions in their oxides. No data was found for the diffusion of cobalt ions in chromium oxides nor for the diffusion of chromium ions in cobalt oxides. Hence, the analysis will be limited to the two other configurations *i.e.* Co//Co and Cr//Cr. Moreover, data in Table 3 are extrapolated from higher temperatures (800-900°C) and the cobalt oxide involved is CoO instead of $Co_3O_4$. According to Table 3, cobalt ions diffuse faster in CoO than chromium ions in $Cr_2O_3$, supporting, at least partially, our hypothesis. More effort needs to be made to support this claims. Hence, the fact that the glaze layer is mainly constituted of $Co_3O_4$ may be related to the ability of $Co_3O_4$ oxides to sinter fastly compared to $Cr_2O_3$ oxides. Eq. (18) can be then rewritten as:

$$N_{GL} \propto \frac{\lambda_{crit}^n f r^m}{k_i \max(D_i)} \quad (20)$$

Table 3: Auto-diffusion coefficients (cm²/s) of cobalt and chromium in their own oxides (CoO and Cr2O3) at 600°C. The temperature in brackets corresponds to the temperature from which the extrapolations at 600°C were made [28–30]

| Chemical element | Auto-diffusion coefficient, $D$ |
|---|---|
| Co | $10^{-12}$ (800°C) – $10^{-10}$ (900°C) |
| Cr | $10^{-13}$ (900°C) – $10^{-20}$ (900°C) |

The quasi pure glaze layer composition is now explained thanks to sintering considerations. It is shown that the diffusive properties of the oxides are predominant in the glaze layer formation. The higher the diffusive coefficient and the faster the generation of the glaze layer. The link between the sintering process and the mild wear process is now strengthened by Eq. (14), experimentally found, and by Eq. (20) which is found in a more physical way.

The design of high-temperature wear material could be made by considering the diffusive properties of each alloying element, as well as their respective quantity. Indeed, it is interesting to promote the rapid glaze layer formation since it lowers the friction coefficient and reduces the amount of wear before its formation. Further experiments could be done with the objective of proposing some design rules for wear-resistant high-temperature alloys.

### 3.4. Discussion on the glaze layer wear mechanism

In the light of these claims, the protective wear action of the glaze layer is discussed (Fig. 16). The first hypothesis to explain the protective behavior of the glaze layer is the plastic accommodation which was discussed in section 2.4. This hypothesis is schematically reminded in Fig. 16 (A). As previously proposed by Viat and co-authors [15], the brittle behavior of the fretted interface at low temperature (T<$T_1$) induces a high debris ejection flow whereas the ductile behavior of the interface at high temperature (T>$T_2$) induces a plastic accommodation and then a reduction of the created debris flow towards zero. However, this hypothesis does not allow to fully understand how the energy is dissipated at the interface (the fretting loop is open at high temperature as illustrated in Fig 2 b)) since there is no production of wear debris. In addition, the effective glaze layer is here very thin (less than 1 µm) and further investigations need to be made to understand how the plastic deformation can be entirely absorbed by this thin layer without any damage and debris production.

A second hypothesis to explain the protective behavior of the glaze layer and related to the tribo-sintering process is proposed hereafter (Fig. 16 B). The driving mechanism of the glaze layer formation (CoRL) is the sintering process which is believed in the second hypothesis to be the key to understand the protective nature of the effective glaze layer [31]. During the steady-state regime, it is proposed that the glaze layer is actually worn by the fretting solicitation and that the tribo-sintering process continuously occurs between the newly created debris particles and the sintered layer. In addition, the quantity of debris produced should be lower than at low temperature since the layer is more ductile, harder and nanocristalline. The wear debris are continuously reincorporated in the glaze layer and therefore the ejected oxide debris flow $\Phi_E$ is drastically limited. From a macroscopic point of view, the wear volume tends towards zero since all the created debris particles are reincorporated and no particles are ejected out of the interface. This hypothesis B is more consistent with the fact that energy is still dissipated in the interface but in a lower quantity than for the low temperature domain (observed through the diminution of the friction coefficient in Fig. 4).

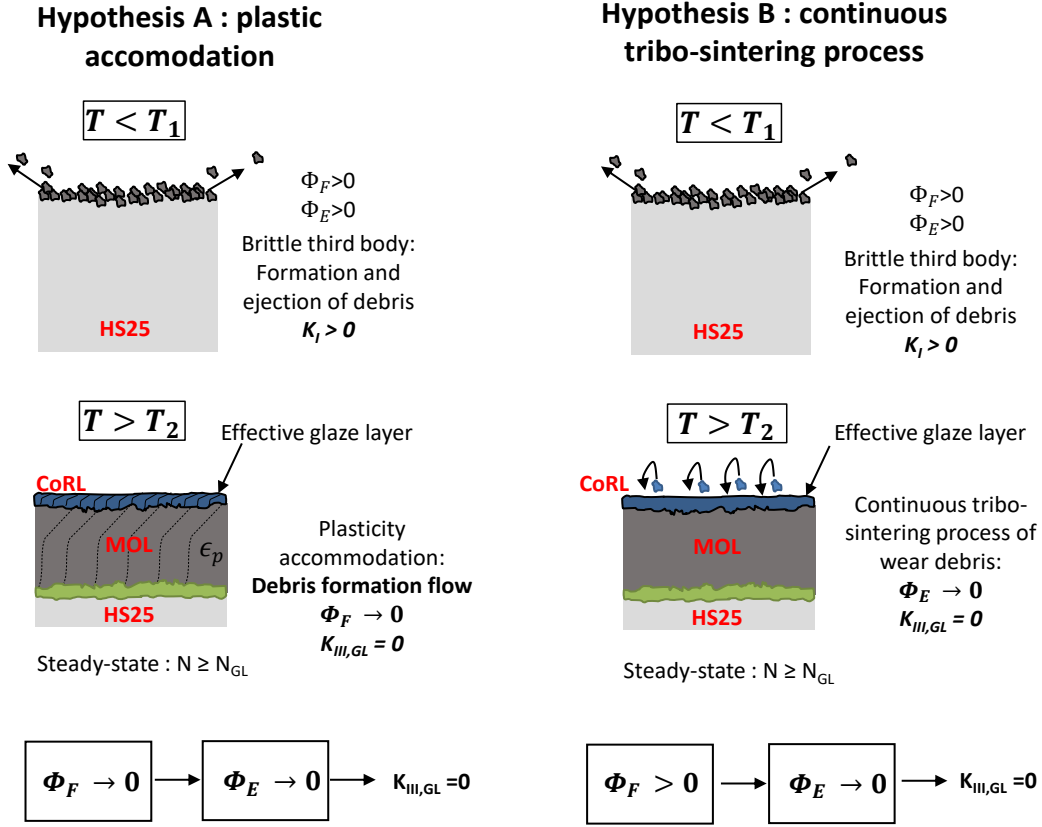

Fig. 16: Protective behavior of the glaze layer through: A) a plastic accommodation process or B) a continuous sintering process of debris particles ($\Phi_F$ reprensents the debris formation flow, $\Phi_E$ represents the debris ejected flow and $K$ is the Archard's wear coefficient)

## 4. Quantitative wear formulation

The formation of the glaze layer is now well explained by performing several microscopic and nanoscopic observations and by formalizing its creation through a tribo-sintering concept. The capacity of the glaze layer to withstand wear is probably due to its capacity to continuously re-incorporate the created wear debris thanks to tribo-sintering. The last section of this paper exposes a formalization of the wear evolution at high temperature.

### 4.1. Archard wear formulation

#### 4.1.1. Effective friction work

A first approach to quantify the wear variations consists in considering the Archard wear formulation. The latter relates the wear volume *V* to the Archard's work defined as the product of the normal force *P* multiplied by the sliding distance *L*:

$$V = K \times \sum W \qquad (21)$$

With *K* the Archard wear coefficient and $\sum W$ the Archard's work equals to:

$$\sum W = PL = 4PN\delta_0 \qquad (22)$$

The Archard approach assumes a linear increase in the wear volume with the fretting cycles, but the previous analyses suggest a bi-linear evolution when the glaze layer is formed (Fig. 5). Therefore, an effective friction work $\Sigma W_{eff}$ is introduced in order to capture the wear kinetics in domains II and III. The main idea is to consider a wear formulation before the glaze layer formation, when $N < N_{GL}$ cycles and to stop calculating wear when $N \geq N_{GL}$ (Fig. 17). The assumption of the absence of wear in domain II and III when the compacted tribolayer is formed is corroborated by the Archard's wear coefficients calculated in section 2 ($K_{II,GL}$ = 2.06.10$^{-6}$ mm$^3$/J, $K_{III,GL}$ = 2.38.10$^{-8}$ mm$^3$/J) which are very low. The friction work is then the sum of the friction work dissipated in the interface before the glaze layer is formed. In order to unify the wear formulation for all the temperatures, an effective number of fretting cycles $N_{eff}$ is introduced [19]:

$$N_{eff} = \begin{cases} N \text{ if } N < N_{GL} \\ N_{GL} \text{ if } N \geq N_{GL} \end{cases} \quad (23)$$

And then,

$$\Sigma W_{eff} = \sum_{N=0}^{N_{eff}} W \quad (24)$$

This wear formulation considers that the wear process when $N < N_{GL}$ at high temperature is the same as that at low temperature, i.e. an oxidative-abrasive wear process.

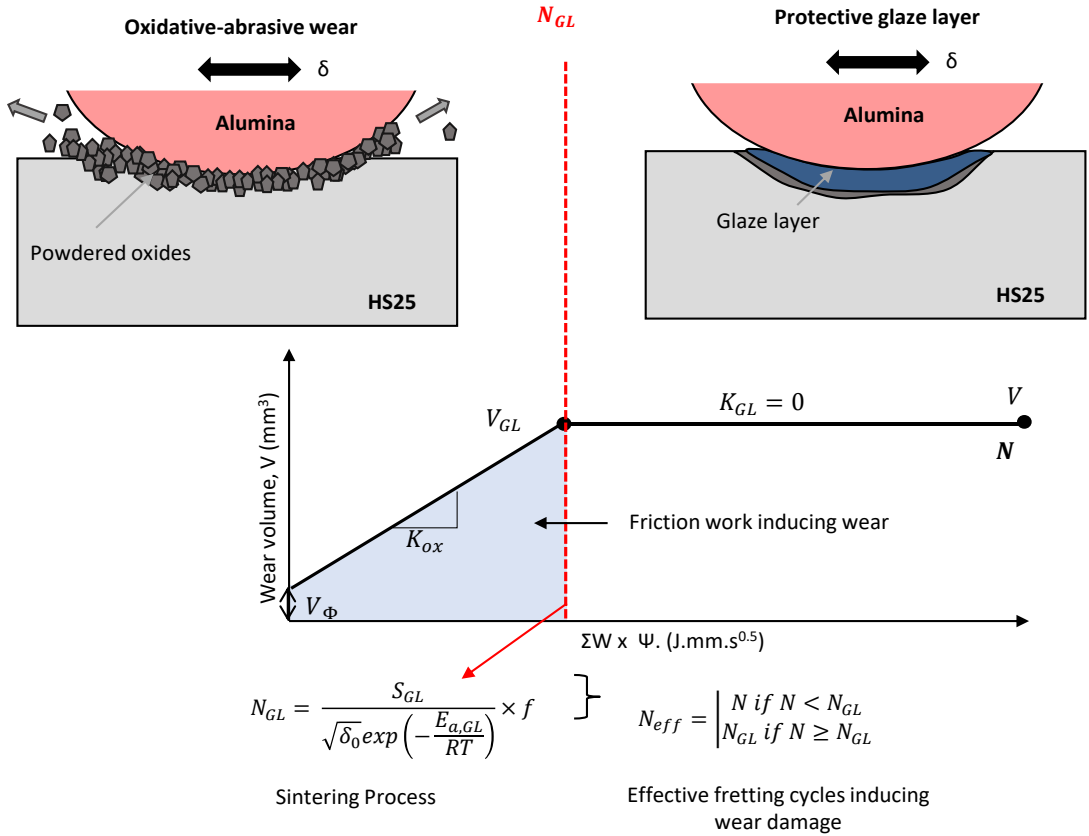

Fig. 17: Introduction of the effective friction work concept in order to describe the wear evolution over a large range of temperatures (severe or mild wear)

*4.1.2. Oxidative-abrasive wear model*

In a previous paper [18], it was found that in domain I (when T < $T_1$), the wear volume depends both on mechanical loadings and chemical reactions. The proposed scenario suggests that wear is controlled by a synergistic interaction of the oxidation of the alloy and its abrasion by the hard counterbody. This scenario accounts for the increase in wear with temperature by considering that increasing the temperature enhances the growth of the oxide layer at the interface and therefore the amount of wear when the latter is abraded. The wear volume in domain I is then described by the following equation:

$$V = \frac{K_{ox}}{\sqrt{f}} \exp\left(-\frac{E_a}{2RT}\right) \delta_0^2 NP + V_\phi \qquad (25)$$

Where $K_{ox}$ is a fitting parameter (mm/s$^{0.5}$/N), also called "oxidational wear coefficient", $f$ is the frequency (Hz), $E_a$ is the activation energy of the oxidation process (J/mol), $R$ is the universal constant, $T$ is the operating temperature (K), $\delta_0$ is the sliding amplitude (mm), $N$ is the number of fretting cycles and $P$ is the normal force. $V_\Phi$ is an offset of wear (mm$^3$) which may happen at the very beginning of the test if the initial contact pressure is high. There are three unknown parameters in Eq. (25) : $K_{ox}$, $E_a$ and $V_\Phi$.

$V_\Phi$ is calculated by considering the wear kinetics as showed in Fig. 4 b). $K_{ox}$ and $E_a$ need to be adjusted on the curve showing the wear evolution with temperature in domain I (Fig. 4 a)). A more detailed procedure of the adjustment of the unknowns is proposed in [18]. In this study, $V_\Phi$ is equal to 0.0071 mm$^3$, $K_{ox}$ is equal to 1.73.10$^{-2}$ mm/s$_{0.5}$/N and $E_a$ is equal to 36748 J/mol.

*4.1.3. Global wear model*

The wear model proposed in Eq. (25) for domain I is extended to all domains by considering the effective Archard's work.

$$V = \frac{K_{ox}}{\sqrt{f}} \exp\left(-\frac{E_a}{2RT}\right) \delta_0^2 N_{eff} P + V_\phi \qquad (26)$$

Where N$_{eff}$ is calculated with Eq. (14) and Eq. (23). However, for each domain $V_\varphi$ and $K_{ox}$ need to be calculated since these parameters seem to be dependent on the operating temperature [19]. Table 4 summarizes the numerical values of the parameters.

Table 4: Numerical values of various parameters for Eq. (26)

| Domain | Temperature range | Wear offset, $V_\phi$ (mm$^3$) | Oxidational wear coefficient, $K_{ox}$ (mm/s$^{0.5}$/N) |
|---|---|---|---|
| I | 25°C ≤ T ≤ 150°C | 0.0071 | 1.73.10$^{-2}$ |
| II | 150°C < T < 300°C | 0.0013 | 3.98.10$^{-3}$ |
| III | 300° ≤ T ≤ 600°C | 0 | 4.75.10$^{-4}$ |

Fig. 18 presents the results of the global wear model applied for the temperature variation and the wear kinetics experimentally performed in Fig. 4. In Fig. 18 a) a good description of the experimental variations is observed for an operating temperature ranging from ambient temperature to 600°C. The model is then able to describe two phenomena: the increase in wear when the temperature increases due to an enhanced oxidation (I); and the decrease in wear at high temperature due to an enhanced tribo-sintering of the wear debris particles (II and III) leading to the formation of the glaze layer. The model is also able to well describe the wear kinetics for various temperatures as showed in Fig. 18 b). At 100°C, the wear increase is linear since the $N_{GL}$ value is not reached after 300 000 cycles. On the contrary, the $N_{GL}$ value is reached at 230°C and 575°C leading to a linear evolution of wear when $N < N_{GL}$ and a constant value of the wear volume when $N \geq N_{GL}$.

This extended Archard's formulation is then able to predict the wear evolution when the temperature and the number of fretting cycles vary but also over a wide range of tribological parameters (frequency, normal load, sliding amplitude and fretting cycles) as previously showed in [19]. One advantage of this global Archard's formulation of wear is that many combinations of parameters can be tested (the friction coefficient is not an input parameter). Hence, Fig. 19 shows the wear evolution with the temperature for different frequencies (5 Hz, 10Hz, 50 Hz and 500 Hz). The model was calibrated using the experiments presented above and performed at 50 Hz.

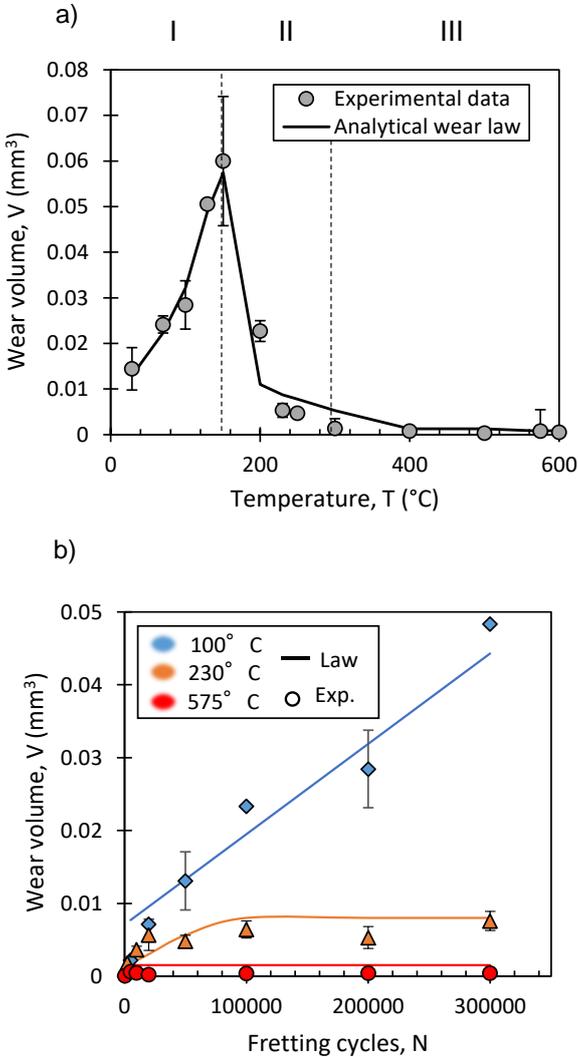

Fig. 18: a) Evolution of the wear volume (experimental and analytical data) when the temperature varies from ambient temperature to 600°C (f = 50 Hz, P = 50 N, δ = ± 20 μm, N = 200 000); b) Wear kinetics (experimental and analytical data) at 100°C, 230°C and 575°C (f = 50 Hz, P = 50 N, $δ_0$ = ± 20 μm)

It is interesting to see that the transition temperature $T_1$, which separates the severe wear domain from the mild wear domain, is not constant. It depends on the frequency: the lower is the frequency, the lower is the transition temperature $T_1$. This is due to the fact that $N_{GL}$ is directly proportional to the frequency and a lower temperature is then needed for a given number of fretting cycles (N = 200 000) to reach $N_{GL}$. This analytical observation was supported by Jin *et al.* [32] who showed that on a stainless steel homogeneous contact, the critical temperature at 20 Hz is around 125°C (temperature at which the wear is maximal) but disappears at 200 Hz. A decrease in the maximum wear volume achieved for the critical temperature is observed analytically in Fig. 19 at 500 Hz. Here, the main explanation of this phenomenon is that the high frequencies reduce the wear volume in domain I (severe wear at low temperature) according to Eq. (25). However, the severe-to-mild transition is still present and the protective tribolayer appears only when $T>T_1$. Jin *et al.* [32] confirms that the wear scars exhibit abrasive wear at low temperature and a compacted oxide layer at high temperature for both tested frequencies.

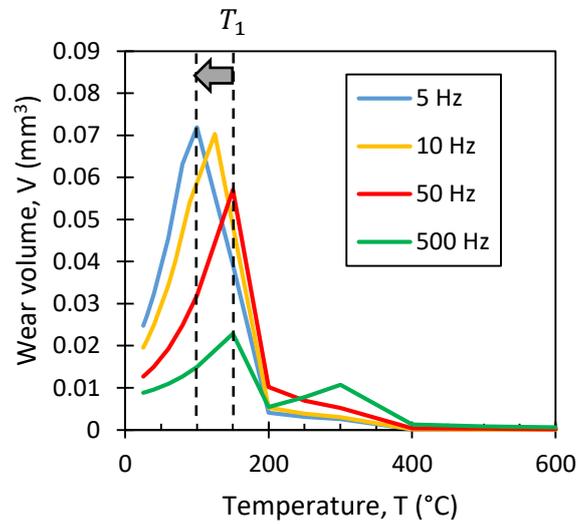

Fig. 19: Theoretical evolution (Eq. (26)) of wear volume as a function of the temperature for different frequencies: 5 Hz, 10 Hz, 50 Hz and 500Hz (P = 50 N, $\delta_0$= ± 20 µm, N = 200 000)

The main advantage of this Archard's formulation is its simplicity and the fact that it considers simple variables like the normal force and sliding amplitude which can be easily estimated in industrial applications using, for instance, FEM computations. However, in this formulation the friction coefficient which is a key aspect in tribological processes [33] is not included. Indeed, the elastic-plastic response of the interface depends on the interfacial shear and therefore on the friction coefficient. To take this aspect into consideration, a friction energy wear approach is presented hereafter.

### 4.2. Energetic wear formulation

As previously proposed [19], the global wear model presented in Eq. (26) can also be written in an energetic form including the variation of the friction coefficient. The friction energy wear approach consists in expressing the wear volume as a function of the cumulated friction energy dissipated in the interface ($\sum E_d$):

$$V = \alpha \sum E_d \qquad (27)$$

With $\alpha$ the energetic wear coefficient. The cumulated friction energy can be approximated for constant conditions by Eq. (28).

$$\sum E_d = 4\delta_0 P N \mu \quad (28)$$

Once again, for the studied conditions, an effective friction energy approach can be considered.

### 4.2.1. Effective friction energy approach

As proposed in [19], Eq. (25) can be written in an energetic formulation thanks to the introduction of an effective dissipated friction energy as proposed in Eq. (29):

$$\sum E_{d,eff} = \sum_{N=0}^{N_{eff}} E_d \quad (29)$$

Then,

$$V = \alpha_{ox} \Psi \sum E_{d,eff} + V_\phi \quad (30)$$

Where $\Psi$ is a tribo-oxidation parameter (mm.s$^{0.5}$) defined as follows:

$$\Psi = \frac{1}{\sqrt{f}} \exp\left(-\frac{E_a}{2RT}\right) \delta_0 \quad (31)$$

And where $\alpha_{ox}$ is the oxidational energetic wear coefficient (mm²/s$^{0.5}$/J or mm/s$^{0.5}$/N) defined as follows:

$$\alpha_{ox} = \frac{K_{ox}}{4\overline{\mu_e}} \quad (32)$$

Where $\overline{\mu_e}$ is the mean energetic friction coefficient. The oxidational wear coefficient $\alpha_{ox}$ has different values depending on the wear domain (I, II, III) as shown in Table 5.

Table 5: Numerical values of the oxidational wear coefficient $\alpha_{ox}$

| Domain | Temperature range | Oxidational wear coefficient, $\alpha_{ox}$ (mm²/s$^{0.5}$/J) |
|---|---|---|
| I | 25°C ≤ T ≤ 150°C | 9.78 |
| II | 150°C < T < 300°C | 1.4 |
| III | 300° ≤ T ≤ 600°C | 6.47.10$^{-2}$ |

By considering Eq. (30), an apparent energetic wear coefficient $\alpha^*$ (mm³/J) is introduced:

$$\alpha^* = \alpha_{ox} \Psi \quad (33)$$

Eq. (30) becomes:

$$V = \alpha^* \sum E_{d,eff} + V_\phi \quad (34)$$

The energetic wear relationship presented in Eq. (34) is equivalent to the classical energetic wear law [34] except that an additional term $\Psi$ is present in the formulation of $\alpha^*$ to take into account the reactivity of the interface in its environment. The wear model presented in Eqs. (30)-(34), originally introduced in [19], shows a very good correlation with experiments (R² = 0.97) over a large range of tribological parameters. One interesting aspect in the formalism presented here is the introduction of the intrinsic oxidational energetic wear coefficient $\alpha_{ox}$ which does not vary with the frequency, the temperature or the sliding amplitude for a given wear domain: I, II or III. On the contrary, $\alpha^*$ varies with the tribological parameters since it depends on $\Psi$ (Eq. (33)). Fig. 20 displays the evolution of $\alpha^*$ with the variation of the frequency, the temperature and the sliding amplitude (only for domain I). Analytical data are compared to experimental data where the apparent energetic wear coefficient is simply calculated (in domain I) by dividing the wear volume by the cumulated dissipated friction energy. Fig. 20 confirms that Eq. (33) is able to describe the variation of the apparent energetic wear coefficient and thus confirms the relevance of the proposed energetic wear model.

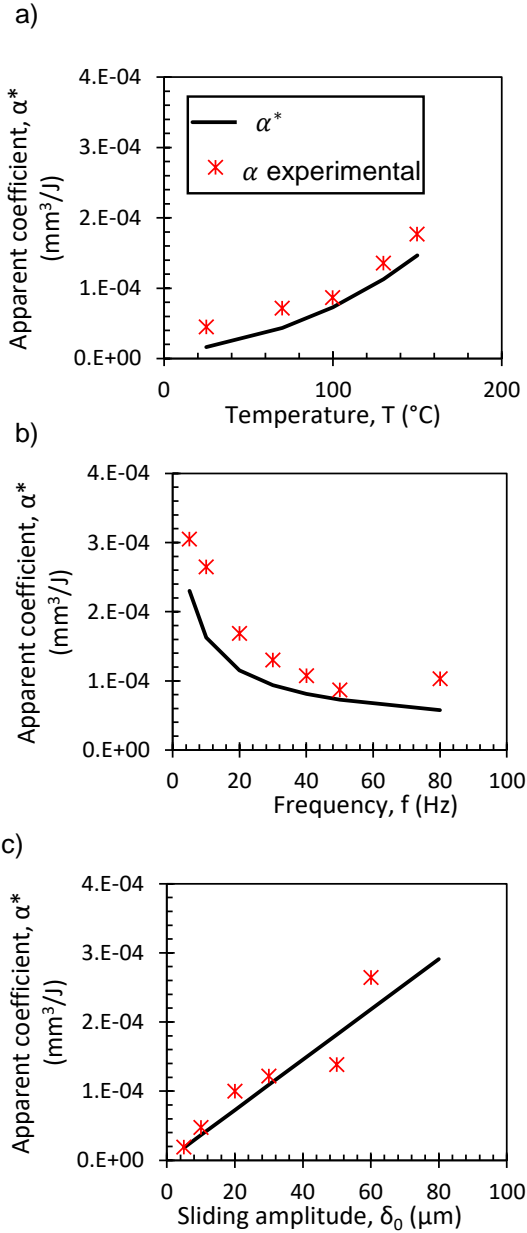

Fig. 20: Evolution of the apparent energetic wear coefficient α* as a function of the temperature (a), the frequency (b) and the sliding amplitude (c) in domain I ($f_{ref}$ = 50 Hz, $P_{ref}$ = 50 N, $\delta_{0,ref}$ = ± 20 μm, N = 200 000)

The dependence of the apparent energetic wear coefficient $\alpha^*$ on the frequency, the sliding amplitude, the temperature is observed by some authors [1,35,36], although the physical meaning of this dependence is not fully understood. This research work attempts to explain the fact that the energetic wear efficiency (and then the apparent energetic wear coefficient) are related to the tribo-oxidation kinetics at the interface through Eq. (31). Hence, the variation of $\alpha^*$ is apparent in the sense that it is possible to decouple this parameter into a constant term ($\alpha_{ox}$) and a variable term ($\Psi$).

*4.2.2. Weighted formulation*

Former research works on fretting wear introduced a weighted friction energy formulation. In these works, a reference test condition was set to define an $\alpha_{ref}$ friction energy coefficient. Then, the influence of loading parameters like the frequency $f$ [35,36], the sliding amplitude $\delta_0$ [36,37], the normal load $P$ [36] and even the contact size $a$ [38] were expressed using the power law formulation below.

$$V = \alpha_{ref} \prod \left(\frac{X}{X_{ref}}\right)^{n_X} \Sigma E_d \qquad (35)$$

$$V = \alpha_{ref} \left(\frac{f}{f_{ref}}\right)^{n_f} \left(\frac{\delta_0}{\delta_{0,ref}}\right)^{n_{\delta_0}} \left(\frac{P}{P_{ref}}\right)^{n_P} \left(\frac{a}{a_{ref}}\right)^{n_a} \Sigma E_d \qquad (36)$$

The reference values ($f_{ref}, \delta_{0,ref}, P_{ref}, \alpha_{ref}$) correspond to the reference test conditions whereas the exponent ($n_f, n_{\delta_0}, n_P, n_a$) monitor the influence of the corresponding variables regarding the energetic wear rate. For instance, in [39,40], the authors showed that the sliding amplitude has a non-linear effect on the wear volume with $n_{\delta_0} = 1$, which implies that $V \propto \delta_0^2$. This suggests that the debris flow is increased with the sliding amplitude inducing a proportional rising of the energetic wear coefficient with the applied sliding amplitude. Dreano *et al.* [18] suggested that the wear rate for the abrasive mechanism is related to the thickness of the oxide layer generated between two sliding strokes using a parabolic law for the oxide growth. This may explain the negative value of $n_f$ found to be equal to -0.5 for stainless steel fretted in corrosive media [41,42] or equal to -0.3 for low-alloy steel fretted in ambient air condition [36]. Finally, Fouvry *et al.* [38] underlined an asymptotic decrease in the energetic wear coefficient with the contact size, leading to a negative exponent $n_a$ = -0.9.

The extended power law formulation introduced in [36] can be considered to capture the effect of the temperature (in domain I, for T<$T_1$). According to Eq. (31), $\Psi$ can be formulated as follows:

$$\Psi = \prod X^{n_X} \qquad (37)$$

Where $X$ is the frequency, the Arrhenius form of the temperature or the sliding amplitude and $n_X$ the associated exponent ($n_f$ = -0.5, $n_{\delta_0}$ = 1 and $n_T$ = 1). A term $\Psi_{ref}$ can be added in Eq. (30):

$$V = \alpha_{ox} \Psi_{ref} \frac{\Psi}{\Psi_{ref}} \Sigma E_d \qquad (38)$$

Thus, Eq. (30) is equivalent to:

$$V = \alpha_{ref} \left\{ \left(\frac{\delta_0}{\delta_{0,ref}}\right) \left(\frac{f}{f_{ref}}\right)^{-0.5} \left(\frac{exp(-E_a/(2RT))}{exp(-E_a/(2RT_{ref}))}\right) \right\} \sum E_d \qquad (39)$$

With the energetic wear coefficient $\alpha_{ref}$ equal to:

$$\alpha_{ref} = \alpha_{ox}\Psi_{ref} \qquad (40)$$

Eq. (39) strengthens the physical meaning of the frequency exponent $n_f$ since the value of -0.5 comes from the tribo-oxidation considerations. Moreover, the influence of the temperature, in the form of an Arrhenius equation, is now predictable through the weighted formulation. In addition, to better interpret the physical meaning of weighted energetic wear formulation, the tribo-oxidation friction energetic analysis demonstrates that the $\alpha_{ref}$ energetic wear coefficient can be expressed as the product of the oxidational energetic wear coefficient $\alpha_{ox}$ multiplied by a $\Psi_{ref}$ which itself expresses the tribo-oxidation condition related to the reference test. Finally, the demonstration below asserts that the global formulation developed in the present paper (Eq. (34)) is equivalent to the weighted formulation developed by other authors in the literature.

5. Conclusion

The paper focused on the wear mechanism of a cobalt-based alloy, enriched with chromium, subjected to fretting at high temperature. The wear volume is very low at high temperature due to the formation of a protective layer (glaze layer) at the interface which is able to resist fretting wear.

Based on previous observations [21], the tribolayer is comprised of three (oxidized) layers in which the chemical composition and the grain size vary. It appeared that the effective glaze layer, i.e. the layer able to resist fretting-wear, is the thin (less than 1µm-thick) cobalt oxide layer at the top of the interface. This layer was suggested to be formed by a tribo-sintering of the compacted debris layer. In order to have a better understanding of the fretted interface, the mechanical properties were extracted thanks to micro-pillar compression. At high temperature, it seems that the fretted interface is more ductile than at low temperature.

Moreover, the tribo-sintering process was investigated by performing fretting-wear on pure metal samples (Cr, Co). It confirmed that the presence of cobalt elements in this alloy is necessary to form a glaze layer (for a given number of cycles). Finally, the variations of the number $N_{GL}$, defined as the number of fretting cycles needed to form a glaze layer, with different tribological parameters are formalized through a sintering concept. It is suggested that cobalt has better diffusive properties in its oxide than chromium and that may explain the formation of the effective glaze layer which is made of cobalt oxide. Finally, it is proposed that the exceptional properties of the glaze layer are due to a continuous re-sintering of the debris particles generated by the friction at the interface. The recirculating process of the wear debris generated from the top cobalt-rich layer prevents wear debris to be ejected from the contact and it therefore drastically reduces wear from a macroscopic point of view.

Finally, a wear model is established, based on previous studies [18,19] and on the tribo-sintering formalization developed in the present paper. The wear model clearly shows a good agreement with the experiment and it is able to describe the wear evolution irrespective of the operating temperature. It is also demonstrated that this wear model provides a physical support for the former semi-empirical weighted energetic wear formulations [35,36,40] which hence strengthens the methodology used in the present research work. The model is validated for a cobalt-based alloy/ceramic contact but can be probably adapted to various metallic contact (steel, Ni-based) or even for pre-formed oxide scale contact. However, this assumption need to be demonstrated, as well as the use of other tribological configurations (cylinder-on-place, flat-on-flat) which can promote a better entrapment of debris inside the interface, and then, promote the glaze layer formation.

Acknowledgments

A part of this work was supported by the LABEX MANUTECH-SISE (ANR-10- LABX-0075) of Université de Lyon, within the program "Investissements d'Avenir" (ANR-11-IDEX-0007) operated by the French National Research Agency (ANR) and by l'EQUIPEX MANUTECH-USD (ANR-10-EQPX-36-01).